\documentclass[10pt,a4paper]{article} % 10pt is ignored!
\pdfoutput=1
\usepackage{jcappub}

\allowdisplaybreaks

\usepackage{hyperref}
\usepackage{ifthen}
\usepackage{amsmath}
\usepackage{amssymb}
\usepackage{color}
\usepackage{graphicx}

\newcommand*{\VEC}[1]  {\textbf{#1}}
\newcommand*{\pp}  {\parallel}
\newcommand*{\df}  {\delta}
\newcommand*{\tf}  {\theta}
\newcommand*{\la}  {\left\langle}
\newcommand*{\ra}  {\right\rangle}

\begin{document}
\notoc  % no contents!

\title{Distribution function approach to redshift space
  distortions. Part V: perturbation theory applied to dark matter halos}
\author[a]{Zvonimir Vlah,}
\emailAdd{zvlah@physik.uzh.ch}

\author[a,b,c,d]{Uro\v{s} Seljak,} 
\emailAdd{seljak@physik.uzh.ch}

\author[d]{Teppei Okumura,} 
\emailAdd{teppei@ewha.ac.kr}

\author[e]{Vincent Desjacques}
\emailAdd{Vincent.Desjacques@unige.ch}

\affiliation[a]{Institute for Theoretical Physics, University of Z\"{u}rich, Z\"{u}rich, Switzerland}
\affiliation[b]{Department of Physics, Department of Astronomy, University of California, Berkeley, CA, USA} 
\affiliation[c]{Lawrence Berkeley National Laboratory, Berkeley, CA, USA}
\affiliation[d]{Institute for the Early Universe, Ewha Womans University, Seoul, S. Korea}
\affiliation[e]{D\' epartement de Physique Th\' eorique and Center for Astroparticle Physics (CAP) Universit\' e de Gen\' eve, Gen\' eve, Switzerland}

%\date{\today}

%===============================================================%
%===============================================================%

\abstract{
Numerical simulations show that redshift space distortions (RSD) introduce strong scale dependence in the 
power spectra of halos, with ten percent deviations relative to linear theory predictions even 
on relatively large scales ($k<0.1h/Mpc$) 
and even in the absence of satellites (which induce Fingers-of-God, FoG, effects). If unmodeled these 
effects prevent one from extracting cosmological information from RSD surveys. In this paper
we use Eulerian perturbation theory (PT) and Eulerian halo biasing model and apply it to the 
  distribution function approach to RSD, in which RSD is decomposed into several correlators of 
density weighted velocity moments. We model each of these correlators using PT and 
compare the results to simulations over a wide range of halo masses and 
redshifts. 
  We find that with an introduction of a physically motivated halo biasing, and using dark matter  
power spectra from simulations,
we can reproduce the simulation results at a percent level on scales up to $k\sim0.15h/Mpc$ at $z=0$, 
without the need to have free FoG parameters in the model. 
}

%===============================================================%


%===============================================================%

\maketitle

%===============================================================%
%===============================================================%
\section{Introduction}
\label{sec:intro}
%===============================================================%
%===============================================================%

Galaxy clustering surveys are one of the most important venues of extracting cosmological information today. 
The reason is that by measuring the 3 dimensional distribution of galaxies we can in principle relate it to the 
3 dimensional distribution of the underlying dark matter, and dark matter distribution is sensitive to many of 
the cosmological parameters. Growth of dark matter structures in time also provides important constraints 
on the models, such as the nature and amount of dark energy in the Universe. Since galaxies are not the perfect tracers of dark matter,
their clustering is biased relative to the dark matter. This means that galaxy surveys cannot determine 
the rate of growth of structure unless this biasing is determined. Fortunately, galaxy redshift surveys provide 
additional information, because the observed redshift is a sum of the radial distance to the galaxy and its peculiar velocity (Doppler shift). 
Galaxies are expected to follow the same gravitational potential as the dark matter and thus are expected to 
have the same velocity (in a large-scale average at least). This leads to a clustering strength that depends on the angle 
between the galaxy pairs and the line of sight, which is referred to as redshift space distortions (RSD). 
In linear theory this can be easily related to the dark matter clustering 
\cite{Kaiser:1987qv,Hamilton:1997zq}. 
These distortions thus make the galaxy clustering in redshift space more complex, but at the same time provide an 
opportunity to extract important information on the dark matter clustering directly from the redshift surveys. 
To what extent this is possible is still a matter of debate: there are significant nonlinear effects that 
spoil this picture, once one goes beyond the very large scales. The goal of this paper is to explore these
nonlinear effects using analytic modeling. 

Velocity induced RSD are usually decomposed into two competing effects: anisotropies along the line of sight due to the 
large coherent motion, so called Kaiser effect, and random motions in virialized objects, known as the
Finger-of-God effect (FoG) \cite{Jackson:2008yv}.  Because of the large coherent motions the RSD anisotropies offer a unique way to measure growth 
rate of structure formation \cite{Cole:1993kh}, and also can provide tests of dark energy models and general relativity 
\cite{White:2008jy,McDonald:2008sh,Yamamoto:2008gr,Reid:2009xm,Percival:2009xn,Kazin:2010nd,Bernstein:2011ju,Reid:2012sw}.  
If good understanding of the nonlinear effects were achieved, RSD would be the most powerful technique 
for these studies because of the fact that redshift surveys provide 3-dimensional information, while other methods, such 
as weak lensing, only provide projected 2-dimensional information (or slightly more if the so-called tomographic 
information is used \cite{Casarini:2012qj}). 

In the past many studies have been performed investigating these nonlinear effects \cite{Scoccimarro:2004tg, Shaw:2008aa, Okumura:2010sv, Taruya:2010mx,
 Jennings:2010uv, Tang:2011qj, Taruya:2013my,Vallinotto:2013eqa}. Some of these methods use analysis and 
modelling based on perturbation theory (for overview see \cite{Bernardeau:2001qr,Carlson:2009it}), but none attempt to 
rely entirely on perturbation theory to explain all of the effects. Instead, they use ansatzes with free parameters, so that if 
the ansatz are accurate one can model the effects accurately. Separately, there have been many approaches
 trying to improve perturbation methods and to increase their ranges of validity 
\cite{Crocce:2005xy, Crocce:2005xz, Crocce:2007dt, Matsubara:2007wj,
  Matsubara:2008wx, McDonald:2006hf, Taruya:2007xy, Pietroni:2008jx,
  Valageas:2003gm, Taruya:2009ir, Taruya:2012ut, Anselmi:2012cn, Crocce:2012fa}. 
Going beyond the dark matter modelling to the dark matter halos and galaxies introduces 
another layer of complication. Dark matter halos are biased relative to the underling dark matter and 
in order to describe them biasing models have to be introduced. Many models have been 
introduced in order to describe this relation \cite{McDonald:2006mx, McDonald:2009dh,
Chan:2012jj,Chan:2012jx,Nishizawa:2012db}, and some have also been included in RSD studies
\cite{Tinker:2006dm, Nishimichi:2011jm, Matsubara:2011ck, 
Matsubara:2012nc, Reid:2011ar, Sato:2011qr,GilMarin:2012nb,Carlson:2012bu,Wang:2013hwa}.

Distribution function approach to modelling the RSD has recently been developed \cite{Seljak:2011tx} 
as a systematic way to study the nonlinear effects in RSD, by decomposing RSD effects into a series of mass-weighted
powers of velocity correlators. This approach has been utilized for the dark matter case 
using the N-body simulations \cite{Okumura:2011pb} and each of the constituent terms has been modeled in PT  
\cite{Vlah:2012ni}.
This model naturally generalizes to the dark matter halos, which have been analyzed using the N-body simulations in \cite{Okumura:2012xh}. 
It has been shown that halo clustering in redshift space has a scale dependence relative to linear theory 
that is stronger than in real space. 
The goal of this paper is to explain this using Eulerian perturbation theory (PT) modelling 
applied to the halos. We adopt the local Eulerian biasing model as a tool to connect the underlying dark matter 
distribution to the halos, but we also explore effects 
beyond the local biasing model, probing the effects of higher order nonlocal terms \cite{Chan:2012jj, Baldauf:2012hs}. 

The paper is organized as follows: we begin in Sec.\ref{sec:model} 
by generalizing the distribution function approach to RSD for dark matter halo case. 
We then introduce the biasing model, and applying it to model the velocity moment correlators. 
In Sec. \ref{sec:ADRMM} we collect all the modelled terms and investigate the total RSD power spectrum results.
We compare the results to the multipoles in Fourier space as well as in configuration space. 
Results are compared to the N-body simulation measurements presented in \cite{Okumura:2012xh}.

For this work, flat $\Lambda$CDM model is assumed 
$\Omega_{\rm m}=0.279$, 
$\Omega_{\Lambda}=0.721$, $\Omega_{\rm b}/\Omega_{\rm m}=0.165$, $h=0.701$,
$n_s=0.96$, $\sigma_8=0.807$. The primordial density field is generated using the matter transfer
function by CAMB. The positions and velocities of all the dark matter halos are given at the redshifts $z=0,~0.509,~0.989$, and 2.070, which are for simplicity
quoted as $z=$0, 0.5, 1, and 2.

%===============================================================%
%===============================================================%
\section{Redshift-space distortions from the distribution function }
\label{sec:model}
%===============================================================%
%===============================================================%

%===============================================================%
\subsection{Definitions and starting equations}
%===============================================================%

Following the recent work on phase space approach to redshift space distortions \cite{Seljak:2011tx,Okumura:2011pb,Okumura:2012xh,
Vlah:2012ni}, distribution function expansion developed for dark matter can be generalized to the dark matter halos.
We can write for halo overdensity field in redshift space
\begin{equation}
     \df^{h}_s(\textbf{k})=\df^{h}(\VEC{k})+\sum_{j=1}\frac{1}{j!}\left(\frac{i k_{\pp}}{\mathcal{H}}\right)^j \mathcal{F}\left[
     \left(1+\df^{h}(\VEC{x})\right)u^j_\pp(\VEC{x})\right](\VEC{k}),
\label{eq:halofield}
\end{equation}
where $u_\pp$ is the halo velocity field projected along the line of sight, and $k_\pp$ projection of the Fourier mode along the line of sight direction.
Also $\mathcal{F}$ stands for the Fourier transformation defined in this paper as;
\begin{align}
    &\tilde{f}(\VEC{k})=\mathcal{F}\left[f(\VEC{x})\right](\VEC{k})=\int{ d^3x ~\text{exp}(i\VEC{k}\cdot\VEC{x})f(\VEC{x})},\nonumber\\
    &f(\VEC{x})=\mathcal{F}^{-1}\left[\tilde{f}(\VEC{k})\right](\VEC{x})\int{\frac{d^3k}{(2\pi)^3}
            ~\text{exp}(-i\VEC{k}\cdot\VEC{x})\tilde{f}(\VEC{k})}.
\end{align}
Using the halo filed expression we can define the dark matter halo-halo power spectrum in redshift-space 
\begin{align}
(2\pi)^3P^{(hh)}_s(\textbf{k})\delta^D(\textbf{k}-\textbf{k}')=\left\langle \delta^h_s(\textbf{k})\delta^{*h}_s(\textbf{k}')\right\rangle.
\end{align}
Dark matter halos are considered to be biased tracers of the underlying dark matter. Identifying the correct biasing model has been proven to 
be a challenging task, and there is still a lot of ongoing work on this subject. 
In the standard local bias approach the halo field can be considered as a functional of the underlying dark matter density 
field $\df^h[\df]$. In
addition to the local relation of the constructed halo field and the underlying matter overdensity field one can also expect 
nonlocal effects (\cite{McDonald:2009dh}). 
For example, in \cite{Chan:2012jj, Chan:2012jx, Baldauf:2012hs} effects of the tidal tensor biasing have been considered. 
In this paper we adopt the model presented in \cite{Baldauf:2012hs}
where the only nonlocal term is due to the second order tidal tensor, which
is added to the standard Taylor expansion of local bias, which we expand to 2nd order.  
We will see that the tidal tensor bias term
does not play the crucial role in modeling the two point statistic of biased object as has already been shown in \cite{Chan:2012jx}, but 
we will nevertheless keep the term in the following expressions. In addition, the third order nonlocal term may also
play an important role in explaining the biasing effects in two point halo statistics studied in this paper. We will include this possibility 
here in a simplified model, deferring a more detailed analysis to a future paper. 
In addition to the nonlocal density 
contributions there are also potential velocity bias effects which, if present, might be important in modelling the halo velocity moments. 
This is a subject of ongoing studies using initial density peaks \cite{Desjacques:2009kt, Desjacques:2010gz}.
Clear results from these studies are still to be determined and so we will not include this possibility here. 
Finally, there are also the effects of exclusion which were emphasized recently in \cite{Baldauf:2013}. 
We will include them in a simplified model. 

We use Eulerian biasing model
\begin{equation}
        \df^h(\VEC{x})=b_1\df(\VEC{x})+\frac{b_2}{2}\left(\df^2(\VEC{x}) -\left\langle\df^2\right\rangle\right)
        +\frac{b_s}{2}\left(s^2(\VEC{x}) -\left\langle s^2\right\rangle\right) +\frac{b_3}{6}\delta^3(\VEC{x}),
\end{equation}
where $\df$ is underlying dark matter overdensity field,  and $b_i$ are the coresponding bias parameters. 
We also add the nonlocal biasing term $b_s$ presented in \citep{Baldauf:2012hs}, and more extensively studied in \citep{Chan:2012jj},
\begin{equation}
        s^2(\VEC{k})=\int{\frac{d^3q}{(2\pi)^3}S^{(2)}(\VEC{q},\VEC{k}-\VEC{q})\df(\VEC{q})\df(\VEC{k}-\VEC{q})},\qquad S^{(2)}(\VEC{k},\VEC{q})=\frac{(\VEC{k}\cdot\VEC{q})^2}{q^2k^2}-\frac{1}{3}.
\label{eq:sker}
\end{equation}
From now on we will use the abbreviations for the integrations of the convolution form
\begin{align}
        (U\circ V)_\VEC{k}=\int{\frac{d^3q}{(2\pi)^3} U(\VEC{q})V(\VEC{k}-\VEC{q})}=U_\VEC{q}V_{\VEC{k}-\VEC{q}}
\end{align}
in order to make following expressions shorter. 
To use this biasing model we need to preform Fourier transformation, which leads to
\begin{equation}
        \df^h_{\VEC{k}}=b_1\df_{\VEC{k}}+\frac{b_2}{2}\df_{\VEC{q}}\df_{\VEC{k}-\VEC{q}}+\frac{b_s}{2}s^2_\VEC{k}+\frac{b_3}{6}\delta_{\VEC{q}}\delta_{\VEC{q}'}
        \df_{\VEC{k}-\VEC{q}-\VEC{q}'},
\label{eq:halofield}
\end{equation}
and for the higher moments of phase space distribution function 
\begin{equation}
        T^h_L=T^{h,\pp}_L(\VEC{k})\equiv\mathcal{F}\left[1/\bar{\rho}\int{d^3q~f_h(\VEC{x},\VEC{q})q_\pp^L}\right],
\end{equation}
where $f_h$ is the phase space distribution function of halos. This gives
\begin{align}
     &\la T^h_L|\right. = \Big[\left(
     1-\sigma^2b_2/2-\sigma_s^2b_s/2\right) \la 1|\right. + b_1 \la
 \df |\right.+\frac{b_2}{2}\la \df^2|\right. +\frac{b_s}{2}\la s^2|\right. +\frac{b_3}{6}\la \df^3|\right. \Big]|\circ u^L_\pp|.
 \label{eq:halofield2}
\end{align}
If we have curl free velocity fields we can write $u^\pp_\VEC{k}=i\frac{k_\pp}{k^2}\theta_\VEC{k}$.

%===============================================================%
\subsection{Halo power spectrum expression}
%===============================================================%

Using expressions \ref{eq:halofield} and \ref{eq:halofield2} for the halo overdensity and higher moments fields 
and definition for the halo power spectrum in redshift space, $P^{(hh)}_{ss}$, we get;
\begin{align}
    P^{(hh)}_{ss,\VEC{k}} &=\sum_{L=0} \sum_{L'=0} \frac{(-1)^{L'}}{L!L'!}\left(\frac{i k_\pp}{\mathcal{H}}\right)^{L+L'}P^{(hh)}_{LL',\VEC{k}}\nonumber\\
    &=\sum_{L=0} \frac{1}{(L!)^2}\left(\frac{k\mu}{\mathcal{H}}\right)^{L+L'}P^{(hh)}_{LL,\VEC{k}}
                  +2Re\sum_{L=0} \sum_{L'>L} \frac{(-1)^{L'}}{L!L'!}\left(\frac{ik\mu}{\mathcal{H}}\right)^{L+L'}P^{(hh)}_{LL',\VEC{k}}.
\end{align}
where $k_\pp/k=\cos\tf=\mu$ and we define 
\begin{align}
(2\pi)^3P^{(hh)}_{LL'}(\VEC{k})\df^D(\VEC{k}-\VEC{k}')=\left\langle T^h_L(\VEC{k})\right.\left|T^{h*}_{L'}(\VEC{k}')\right\rangle. 
\label{eq:psdef}
\end{align}
In this paper we will consider terms that contribute up to one loop in Eulerian PT, so we have the contributing terms
\begin{align}
    P^{(hh)}_{ss,\VEC{k}} =& P^{(hh)}_{00,\VEC{k}}+\left(\frac{k\mu}{\mathcal{H}}\right)^2P^{(hh)}_{11,\VEC{k}}+\frac{1}{4}\left(\frac{k\mu}{\mathcal{H}}\right)^4P^{(hh)}_{22,\VEC{k}}\nonumber\\
    &+2\text{Re}\left[\left(\frac{-ik\mu}{\mathcal{H}}P^{(hh)}_{01,\VEC{k}}\right)
    +\left(-\frac{1}{2}\left(\frac{k\mu}{\mathcal{H}}\right)^2 P^{(hh)}_{02,\VEC{k}}\right)
    +\left(\frac{i}{6}\left(\frac{k\mu}{\mathcal{H}}\right)^3 P^{(hh)}_{03,\VEC{k}}\right)\right.\nonumber\\
    &+\left.\left(-\frac{i}{2}\left(\frac{k\mu}{\mathcal{H}}\right)^3P^{(hh)}_{12,\VEC{k}}\right)
    +\left(-\frac{1}{6}\left(\frac{k\mu}{\mathcal{H}}\right)^4P^{(hh)}_{13,\VEC{k}}\right)
    +\left(\frac{1}{24}\left(\frac{k\mu}{\mathcal{H}}\right)^4P^{(hh)}_{04,\VEC{k}}\right)\right].
\label{eq:galaxypower}
\end{align} 
Strictly speaking $P^{(hh)}_{04}$ is of the higher order, but has been proven to be significant \cite{Vlah:2012ni}, so we are including it in the model.
As it is discussed in \cite{Seljak:2011tx, Vlah:2012ni}, going to higher order in this expression we introduce higher order velocity moments. Since in our biasing model 
we assume no velocity biasing, at 1-loop level higher order terms will give small difference from the linear biasing model from dark matter halo power spectrum. 
Terms in which the nonlinear biasing model plays a substantial difference at 1-loop level treatment are $P^{(hh)}_{00}$, $P^{(hh)}_{01}$, $P^{(hh)}_{02}$, $P^{(hh)}_{11}$.
In the following part of this section we are considering these terms in more detail.

%===============================================================%
\subsection{Halo-matter $P^{(hm)}_{00}$ power spectrum}
%===============================================================%

First we consider the $P^{(h \bar{h})}_{00}$ term, the real space power spectrum. 
Formally we make the distinction between two halo tracers labeling the second bias coefficients with the bar.
This term is isotropic and does not have $\mu$ dependence.
We will first model the cross-correlation with the matter. 
For convenience we use the different biasing coefficients for
autocorrelation of the halo overdensity field,
\begin{align}
     \la \df^{h}|\df^{\bar{h}} \ra 
&= b_1 \bar{b}_1\la\df|\df\ra + \frac{1}{4}b_2
\bar{b}_2\la\df^2|\df^2\ra + \frac{1}{4}b_s \bar{b}_s\la s^2|s^2\ra +\frac{1}{36}b_3 \bar{b}_3 \la\df^3|\df^3\ra\nonumber\\
&+\{b_1,b_2\}\la\df|\df^2\ra + \frac{1}{3}\{b_1,b_3\}\la\df|\df^3\ra + \frac{1}{6}\{b_2,b_3\}\la\df^2|\df^3\ra \nonumber\\
&+\{b_1,b_s\}\la\df |s^2\ra + \frac{1}{2}\{b_2,b_s\}\la\df^2|s^2\ra + \frac{1}{6}\{b_3,b_s\}\la\df^3|s^2\ra
\end{align}
where we have the anticommutator defined as
$\{a,b\}=(a\bar{b}+b\bar{a})/2$. Keeping different biasing coefficients enables us to use this result also for the
matter-halo cross spectrum $P^{(hm)}_{00}$ by choosing $b_1=1$, and $b_2=b_s=b_3=0$ for dark matter. 
For the matter-halo cross spectrum it follows
\begin{equation}
     \la \df^m|\df^h \ra = b_1\la\df|\df\ra +
     \frac{1}{2}b_2\la\df|\df^2\ra  +\frac{1}{2}b_s\la\df |s^2\ra+\frac{1}{6}b_3\la\df|\df^3\ra.
\end{equation}
We restrict our consideration to the 1-loop PT modelling at this level, and use bias renormalization techniques presented in \cite{McDonald:2006hf}.
The same bias renormalization is also used in all the other terms at the 1-loop level. 
At this order the relevant bias terms are 
\begin{align}
    \la\df_\VEC{k}|\df_{\VEC{k}'}\ra &\sim P_{00,\VEC{k}},\nonumber\\
    \la\df_\VEC{k}|\df^2_{\VEC{k}'}\ra &\sim 2F^{(2)}_{\VEC{q},\VEC{k}-\VEC{q}}P^{(L)}_{\VEC{q}}P^{(L)}_{\VEC{k}-\VEC{q}}+2\frac{34}{21}P^{(L)}_\VEC{k}\sigma^2,\nonumber\\
    \la\df_\VEC{k}|\df^3_{\VEC{k}'}\ra &\sim 3\sigma^2 P_{00,\VEC{k}},\nonumber\\
    \la\df_\VEC{k}|s^2_{\VEC{k}'}\ra &\sim 2F^{(2)}_{\VEC{q},\VEC{k}-\VEC{q}}S^{(2)}_{\VEC{q},\VEC{k}-\VEC{q}}P^{(L)}_{\VEC{q}}P^{(L)}_{\VEC{k}-\VEC{q}},
\label{ch2:eq3}
\end{align}
where $\sim$ symbol means that we have dropped $(2\pi)^3\df^D(\VEC{k}-\VEC{k}')$ factor from the left hand site of the relations above.  
From this follows for the cross power spectrum
\begin{equation}
     P^{(hm)}_{00,\VEC{k}}=\left(b_1+b_2\frac{34}{21}\sigma^2+\frac{b_3}{2}\sigma^2\right)P_{00,\VEC{k}}
     + b_2K_{00,\VEC{k}}+b_s K^s_{00,\VEC{k}},
\end{equation}
where we have
\begin{align}
     &K_{00,\VEC{k}}\equiv
     P^{(L)}_{\VEC{q}}P^{(L)}_{\VEC{k}-\VEC{q}}F^{(2)}_{\VEC{q},\VEC{k}-\VEC{q}},\nonumber\\
    &K^s_{00,\VEC{k}}\equiv
     P^{(L)}_{\VEC{q}}P^{(L)}_{\VEC{k}-\VEC{q}}F^{(2)}_{\VEC{q},\VEC{k}-\VEC{q}}S^{(2)}_{\VEC{q},\VEC{k}-\VEC{q}}
\label{eq:k00terms}
\end{align}
where $F^{(2)}$ is the standard overdensity kernel in Eulerian PT (e.g. \cite{Bernardeau:2001qr,Vlah:2012ni}) and $S^{(2)}$ 
kernel is defined in equation \ref{eq:sker}.
Using the bias renormalisation
\begin{align}
     &b_1\rightarrow b'_1=b_1+\frac{34}{21}\sigma^2b_2+\frac{1}{2}\sigma^2b_3,\nonumber\\
     &b_2\rightarrow b'_2=b_2,\nonumber\\
     &b_s\rightarrow b'_s=b_s,
\end{align}
we get
\begin{equation}
     P^{(hm)}_{00,\VEC{k}}=b_1P_{00,\VEC{k}} + b_2K_{00,\VEC{k}}+b_sK^s_{00,\VEC{k}}.
\label{eq:phm}
\end{equation}
For halo-halo power spectrum we can take $\bar b=b$, and then
in addition to the terms in \ref{eq:k00terms} we have 
\begin{align}
    \la\df^2_k|\df^2_{k'}\ra &\sim
    2P^{(L)}_{\VEC{q}}P^{(L)}_{\VEC{k}-\VEC{q}}\equiv 2K_{01,\VEC{k}},\nonumber\\
    \la s^2_k| s^2_{k'}\ra &\sim
    2\left(S^{(2)}_{\VEC{q},\VEC{k}-\VEC{q}}\right)^2P^{(L)}_{\VEC{q}}P^{(L)}_{\VEC{k}-\VEC{k}}\equiv 2K^s_{01,\VEC{k}},\nonumber\\
    \la\df^2_k|s^2_{k'}\ra &\sim
    2S^{(2)}_{\VEC{q},\VEC{k}-\VEC{q}}P^{(L)}_{\VEC{q}}P^{(L)}_{\VEC{k}-\VEC{q}}\equiv 2K^s_{02,\VEC{k}},\nonumber\\
\label{ch2:eq3}
\end{align}
after the renormalization this gives
\begin{equation}
     P^{(hh)}_{00,\VEC{k}}=b_1^2 P_{00,\VEC{k}}+2b_1\Big[ b_2
     K_{00,\VEC{k}}+b_sK^s_{00,\VEC{k}}\Big]+\frac{1}{2}\Big[ b^2_2
     K_{01,\VEC{k}}+ b^2_s K^s_{01,\VEC{k}}\Big]+b_2b_s K^s_{02,\VEC{k}}.
\label{eq:phh}
\end{equation}
This expression still lacks the stochasticity terms coming from the discreteness of the halos, 
such as the Poisson shot noise or its generalizations, discussed further below. 

\begin{figure}[t]
    \centering
    \includegraphics[scale=0.45]{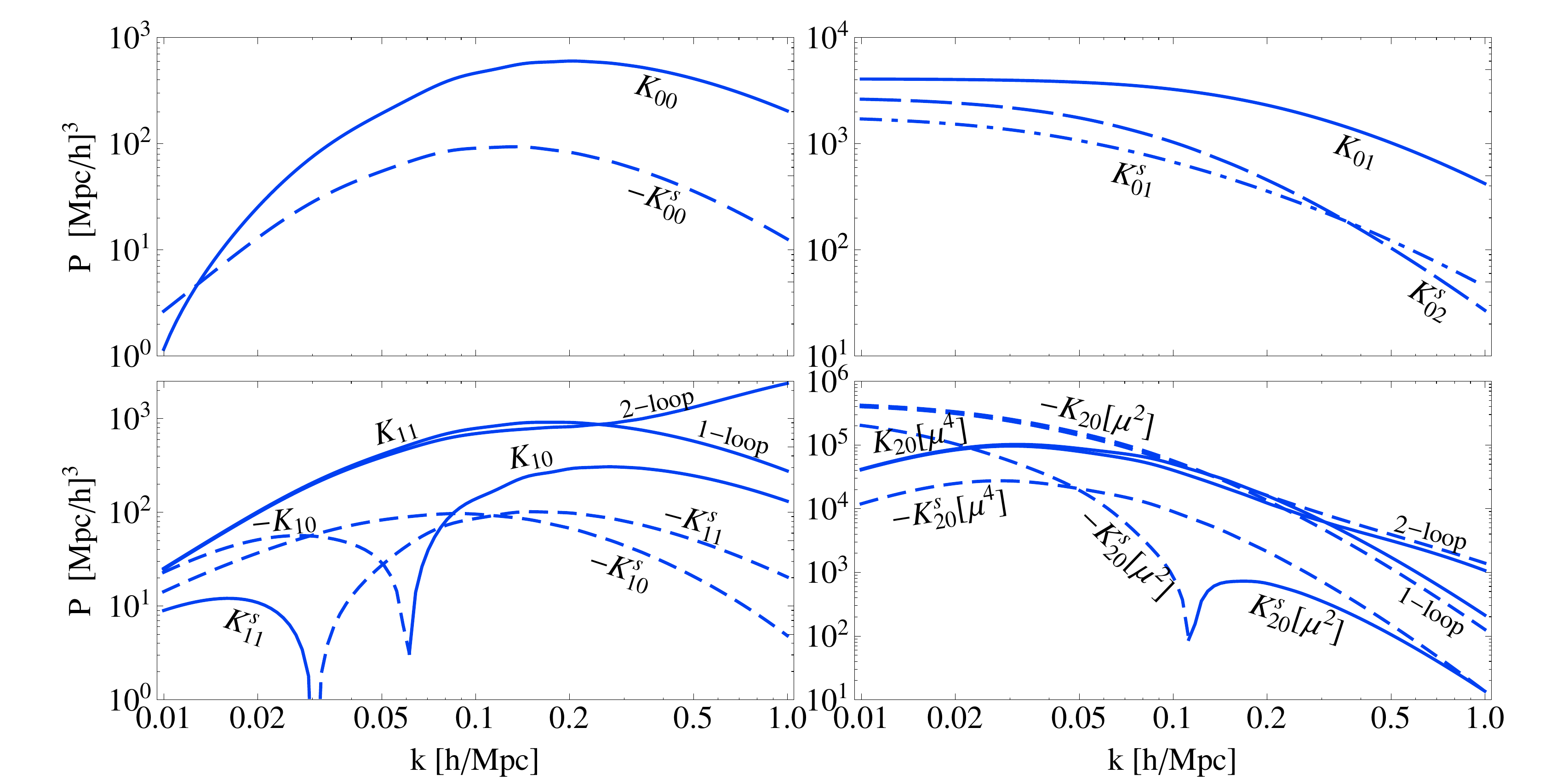}
    \caption{\small $K_{ij}$ terms using one loop PT results, and also two loop results for some of these terms. 
    The terms proportional to $b_s$  are denoted with a subscript $s$ and can be compared to the 
term proportional to $b_2$. 
    Solid lines are for positive values and dashed are for negative.}
    \label{fig:1}
\end{figure}  

%=================================%
\subsubsection{Nonlinear bias terms $K_{ij}$}
\label{sec:bs}
%=================================%

Modelling the biasing of dark matter halos beyond the linear bias $b_1$ introduces additional 1-loop terms in 
perturbative approach. These terms are labeled here with $K_{ij}$ and will show up 
in the first few moment power spectra $P^{hh}_{00}, P^{hh}_{01}, P^{hh}_{02}$, where $b_2$, $b_s$ and $b_{3,nl}$ 
(the nonlocal third order bias \cite{McDonald:2009dh,Chan:2012jx,Saito:2013})
appear at 1-loop level. In figure \ref{fig:1} we show the k-dependence of $K_{ij}$ terms. The relative contribution of these terms depends on the amplitudes of 
$b_2$, $b_s$ and $b_{3,nl}$. For example, one can use the values of $b_2$, $b_s$ from bispectrum \cite{Baldauf:2012hs,Chan:2012jx}, where
$b_s$ terms are small relative to the $b_2$. Taking into account also the $k$-dependence of these terms we can see that effectively $b_s$ effects can be absorbed 
in the renormalized $b_2$ value. For $b_{3,nl}$ the corresponding trispectrum analysis has not been performed yet, but a power 
spectrum analysis together with the coevolution values \cite{Saito:2013} suggests that it can be important relative to $b_2$. 

Over a limited range of interest where these terms matter and we are not too deeply into nonlinear regime, 
$0.05h/Mpc<k<0.2h/Mpc$, these terms can be organised (considering the $k$-dependences and relative amplitudes) in a way that for each statistics we 
can define a single effective nonlinear bias parameter. In case of $P_{00}$ we will denote it $b_2^{00}$, and in case of $P_{01}$, $b_2^{01}$. 

%===============================================================%
\subsubsection{Effective model for halo-halo power spectrum $P^{(hh)}_{00}$}
%===============================================================%

\begin{figure}[t!]
%    \centering
    \raggedright
    \includegraphics[scale=0.44]{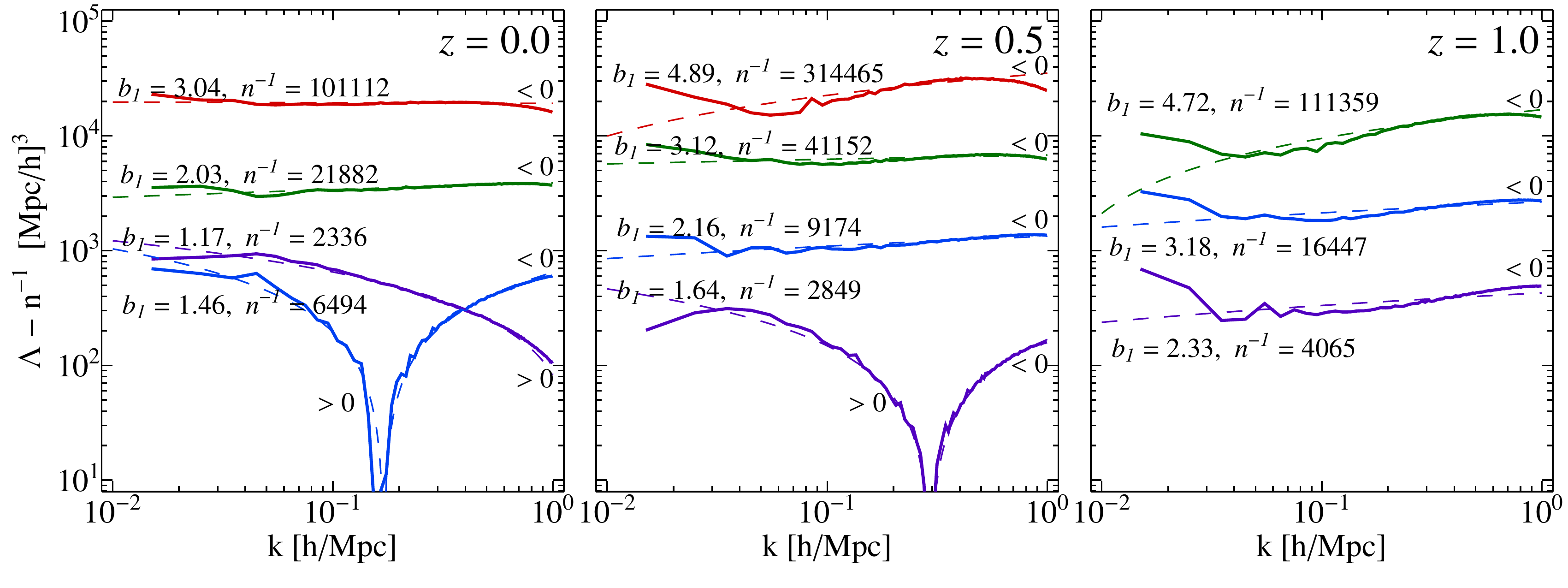}
    \caption{\small The stochasticity as measured from the simulations, once the Poisson shot noise is subtracted (full lines). Four different mass bins 
    are shown at the redshift $z=0.0$ and $z=0.5$ and three at the redshift $z=1.0$. Most of the plotted lines are negative, i.e. stochasticity is sub-Poissonian,  
    except for the lowest mass bins at redshift $z=0.0$ and $z=0.5$ which are positive on large scales. Simple model from equation \ref{eq:SNmodel2} 
    is also shown (dashed lines).
    }
    \label{fig:2}
\end{figure}

Following the arguments from previous section 
we will absorb $b_s$ and $b_{3,nl}$ terms into $b_2$ and omit them from the analysis of $P^{hh}_{00}$ in this section. 
For a complete model of $P^{hh}_{00}$ from $P^{hm}_{00}$ we need to model the stochasticity term, 
\begin{equation}
\Lambda(k)=P^{(hh)}_{00,\VEC{k}}-2b_1P^{(hm)}_{00,\VEC{k}}+b_1^2P^{(mm)}_{00,\VEC{k}}.
\label{eq:phhmodel}
\end{equation}
This term has recently been studied extensively in \cite{Baldauf:2013}, where it is denoted as the diagonal term of the 
stochasticity matrix $C_{ij}$. In the simplest models the stochasticity is given by the Poisson shot noise $\bar{n}^{-1}$, 
where $\bar{n}$ is the halo number density. 
As discussed in \cite{Baldauf:2013}, there are deviations from this prediction 
sourced by both the halo exclusion and the nonlinear clustering of halos relative to dark matter: the latter can be seen from 
upper right panel of figure \ref{fig:1}, where we see that all the terms are constant at low $k$, suggesting a white noise 
contribution at low $k$. Similarly, imposing a finite radius on halos lowers the stochasticity in the low $k$ limit below the 
Poisson $\bar{n}^{-1}$ value. 
In the $k \rightarrow 0$ limit one can determine  from the clustering of halos in initial conditions: the dominant positive 
contribution comes from local biasing of initial halos and the negative contribution comes from exclusion. This limit 
does not change when evolving halos from initial to final redshift. 
The scale dependence of this term however changes from initial to final redshift and the theoretical modeling is still 
poorly understood. In this paper we add this model for the stochastic noise to our PT model of RSD and compare the result to the simulations. 
These shot noise effects are isotropic so the term does not have $\mu$ dependence. It affects only the
modelling of $P^{(hh)}_{00}$ term, while all the higher order velocity moments contributions to RSD power spectra 
are independent of it. To determine it  we use the N-body simulation measurements presented in \cite{Okumura:2011pb, Okumura:2012xh}. 
In figure \ref{fig:2}  we show the results for $\Lambda(k)$ as measured from simulations, subtracting out the  
Poisson shot noise $\bar{n}^{-1}$ (which are given in \cite{Okumura:2012xh}). From the figure we see that
deviations from Poisson model are of order 10-20\% at low $k$ and thus cannot be 
neglected. Most of the plotted lines are negative, i.e. the measured stochasticity
is sub-Poissonian, except for the lowest mass bins at redshift 
$z=0.0$ and $z=0.5$, where it is positive on large scales. As discussed in \cite{Hamaus:2010im, Baldauf:2013} this is because exclusion dominates
over nonlinear biasing 
for higher mass halos, while the opposite is true at low masses. 

To model this $k$ dependence on the scales of interest we propose a simple model
\begin{equation}
\Lambda-\bar{n}^{-1}=\lambda+ p \log k,
\label{eq:SNmodel2}
\end{equation}
where we fit for the values of $\lambda$ and $p$ and show the results in the same figure \ref{fig:2}  as dashed lines. We see that this 
reproduces the measurements over a broad range of $k$ values, specially around $k \sim 0.1h/Mpc$, where its effects are most important.  

\begin{figure}[t!]
    \centering
    \includegraphics[scale=0.45]{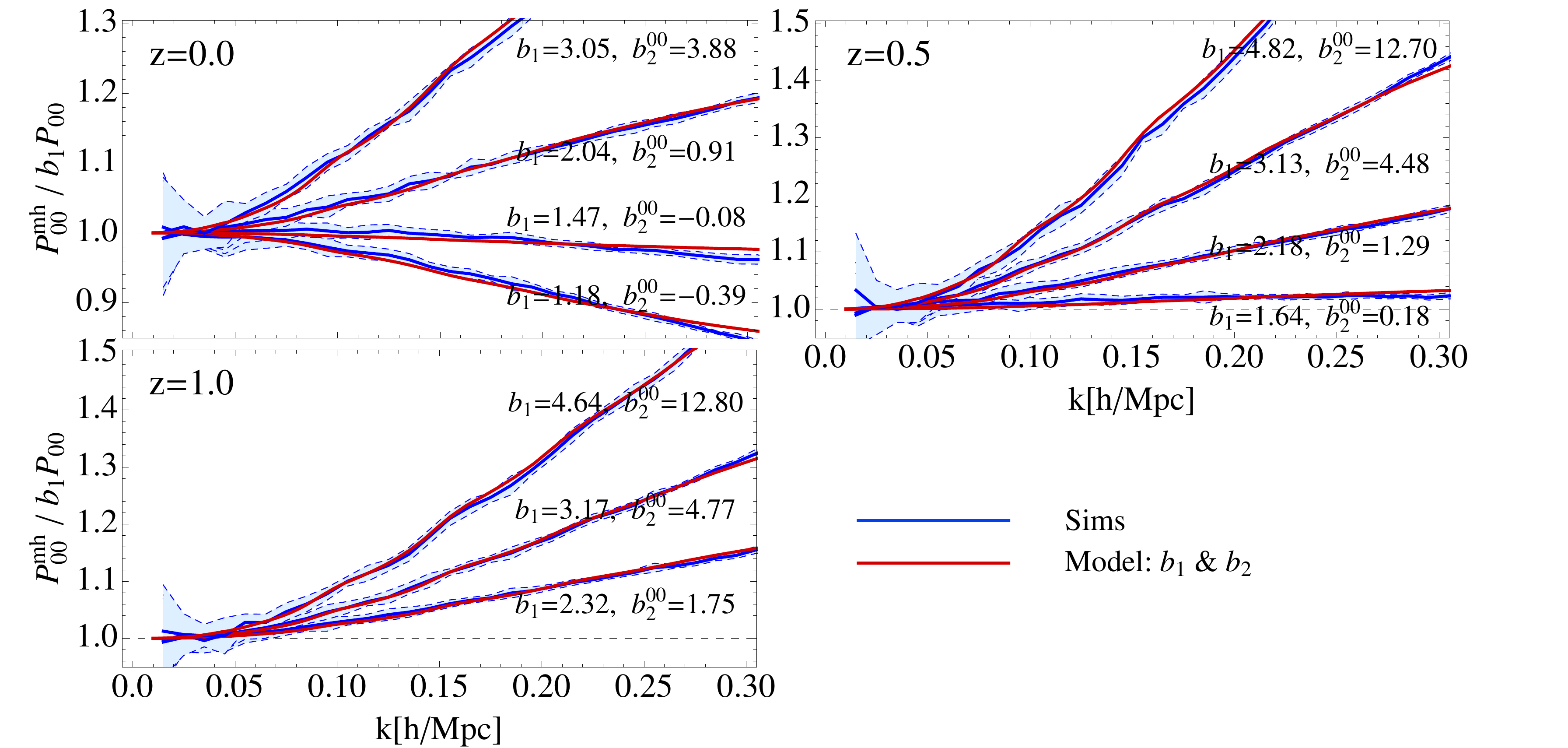}
    \caption{\small Scale dependence of halo-matter cross power spectrum $P^{(hm)}_{00}$ is shown for several mass bins, at reshifts $z=0.0$, $z=0.5$
    and $z=1.0$. Simulation measurements (blue solid line) are shown, as well as the results of the model presented in the text.
    We have fitted for $b_2$ that reproduces best simulation measurements. All the lines are divided by $b_1$ times the 
    dark matter power spectrum from simulations.}
    \label{fig:3}
\end{figure}

\begin{figure}[t!]
    \centering
    \includegraphics[scale=0.44]{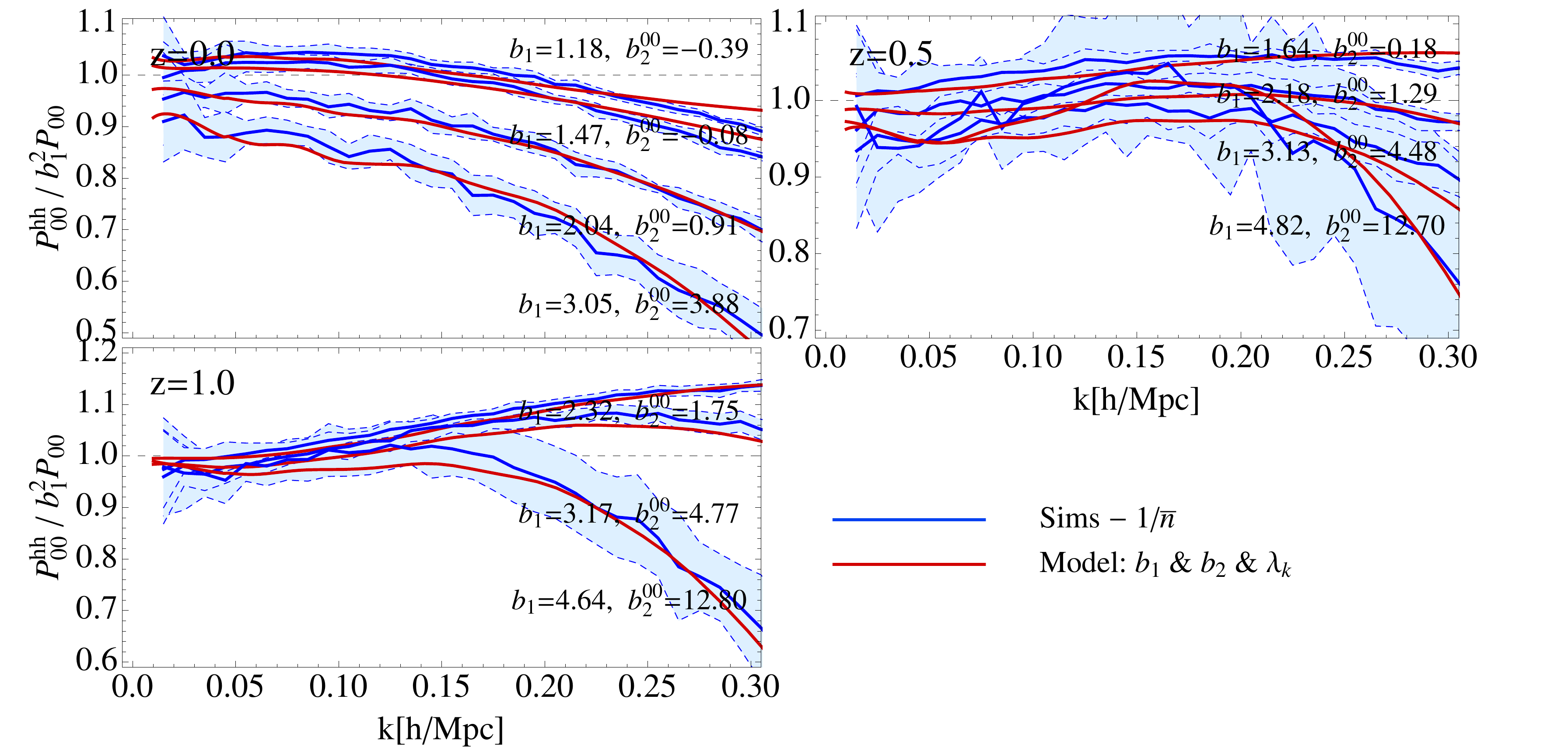}
    \caption{\small Scale dependence of halo-halo auto power spectrum $P^{(hh)}_{00}$ shown for several mass bins, at reshifts $z=0.0$, $z=0.5$
    and $z=1.0$. Simulation measurements (blue solid line) are shown, as well as the results of the model presented in the text. 
    All the lines are divided by the $b_1^2$ times the 
    dark matter power spectrum from the simulations.}
    \label{fig:4}
\end{figure}

In figure \ref{fig:3} we show the scale dependence of the halo-matter cross power spectrum $P^{(hm)}_{00}$ for several mass bins, at redshifts $z=0.0$, $z=0.5$ 
and $z=1.0$.  We fit for an effective $b_2^{00}$ parameter to reproduce simulation measurements on the scales of interest and use these values then to model $P^{(hh)}_{00}$. 
In figure \ref{fig:4} we show the scale dependence of halo-halo auto power spectrum $P^{(hh)}_{00}$ for the same mass bins as before, at reshifts $z=0.0$, $z=0.5$ 
and $z=1.0$. Using the stochasticity model presented above we evaluate it from 
\begin{equation}
P^{(hh)}_{00,\VEC{k}}=2b_1P^{(hm)}_{00,\VEC{k}}-b_1^2P^{(mm)}_{00,\VEC{k}}+\Lambda(k).
\label{eq:phhmdel}
\end{equation}
In this model we also use simulation predictions for dark matter $P^{(mm)}_{00}$ term at each of the redshifts. 

%===============================================================%
\subsection{$P^{(hh)}_{01}$ term}
%===============================================================%

\begin{figure}[t!]
    \centering
    \includegraphics[scale=0.3]{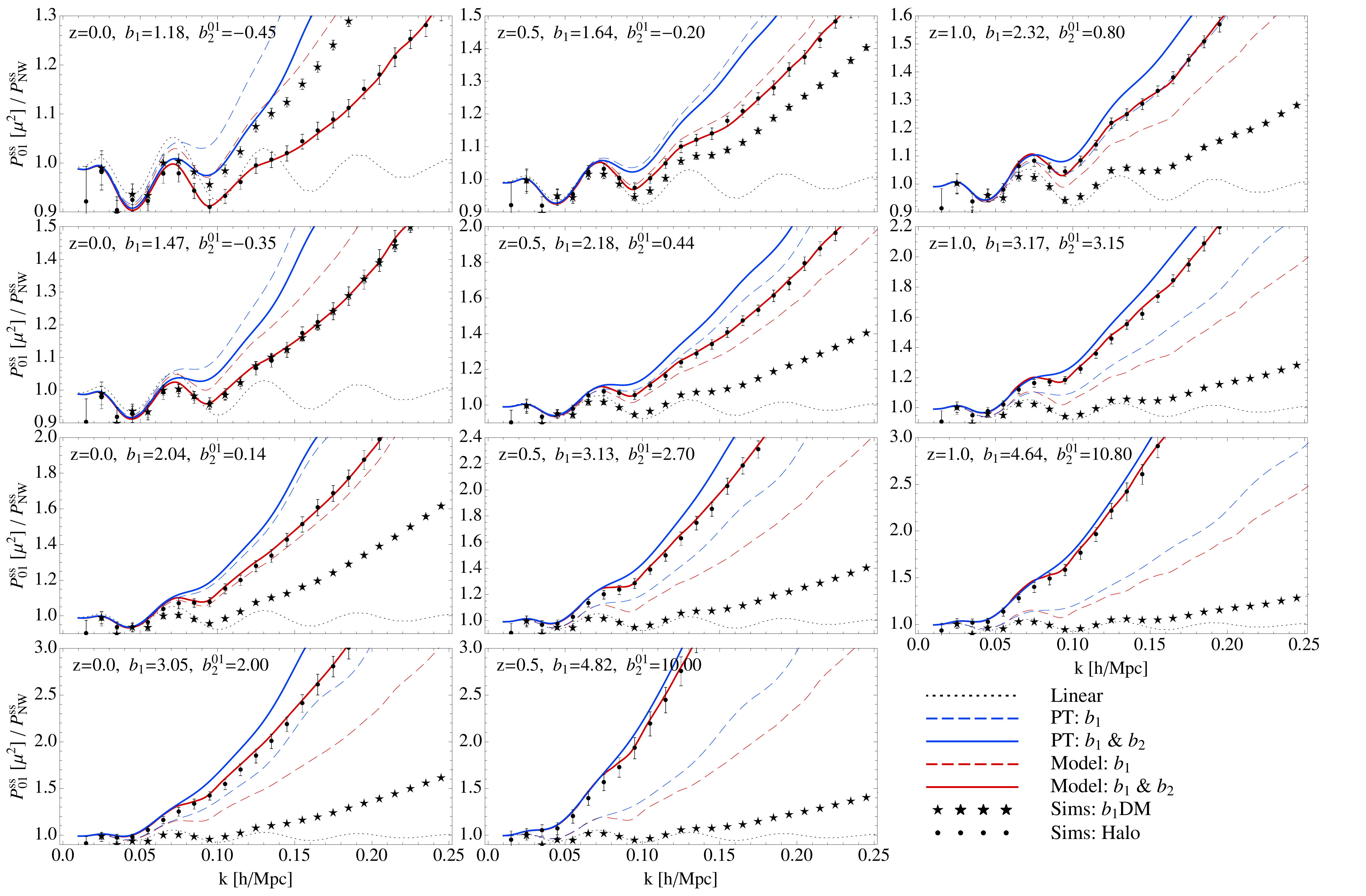}
    \caption{\small Halo-halo density-momentum power spectrum $P_{01}$ for four mass bins, at reshifts $z=0.0$, $z=0.5$
    and $z=1.0$. We show SPT results (blue lines) and, because SPT does not reproduce DM simulations well, we also show
results using the model where we use DM simulations for linear biasing terms (red lines) to isolate the biasing effects in RSD.
    For comparison we also show linear biasing (dashed lines) and nonlinear biasing (solid lines) models. 
Halo (black points) simulation data and $b_1$ times the dark matter data (black stars)
    are also shown. All the spectra are divided by the no wiggle linear prediction \cite{Eisenstein:1997ik} for $b_1P_{01}$ term.}
    \label{fig:5}
\end{figure}

Next we consider a correlation of the halo field and the halo momentum field. This term
has only a scalar contribution and is the leading term in the $\mu^2$ dependence of the total 
redshift power spectrum, as was shown in \cite{Seljak:2011tx, Vlah:2012ni}. 
Here we generalize this terms to halos. Using definitions of the halo density \ref{eq:halofield} and momentum fields \ref{eq:halofield2} we get 
\begin{align}
\la \df^{h} | T_1^{\bar{h}} \ra &=b_1 \la \df|(1+\bar{b}_1\df)u_\pp\ra + \frac{1}{2}\Big[b_2\la \df^2|u_\pp\ra+ b_s \la s^2| u_\pp\ra\Big]
+\frac{1}{2}b_1\bar{b}_2\la \df | \df^2 v_\pp \ra_c\nonumber\\
&+\frac{1}{2}\bar{b}_1\Big[b_2\la \df^2|\df u_\pp\ra+ b_s \la s^2|\df u_\pp\ra \Big] 
+\frac{1}{4}b_2\bar{b_2}\la\df^2|\df^2 v_\pp \ra_c + \frac{1}{6}b_3\la \df^3|u_\pp\ra_c,
\end{align}
where we have used the same renormalization scheme as for $P^{(hh)}_{00}$ term. Subscript $c$ 
stands for the connected part of the correlator, while the disconnected parts get renormalized.
Keeping the terms at one loop order we get for the power spectrum, 
\begin{align}
P^{(h \bar{h})}_{01,\VEC{k}}=b_1\bar{b}_1P_{01,\VEC{k}}+b_1(1-\bar{b}_1)\alpha
P_{\df\tf,\VEC{k}}+
\alpha\Big[b_2K_{10,\VEC{k}}+b_sK^s_{10,\VEC{k}}\Big]+\alpha\bar{b}_1\Big[b_2K_{11,\VEC{k}}+b_sK^s_{11,\VEC{k}}\Big],
\label{eq:P01}
\end{align}
where we have $\alpha=-i\mu/k$. Using the PT to evaluate the contributions of nonlinear biasing terms we get
\begin{align}
K_{10,\VEC{k}}&\equiv
     P^{(L)}_{\VEC{q}}P^{(L)}_{\VEC{k}-\VEC{q}}G^{(2)}_{\VEC{q},\VEC{k}-\VEC{q}},\nonumber\\
K^s_{10,\VEC{k}}&\equiv
     P^{(L)}_{\VEC{q}}P^{(L)}_{\VEC{k}-\VEC{q}}G^{(2)}_{\VEC{q},\VEC{k}-\VEC{q}}S^{(2)}_{\VEC{q},\VEC{k}-\VEC{q}},\nonumber\\
K_{11,\VEC{k}}&\equiv
     \frac{kx}{q}P^{(L)}_{\VEC{q}}P^{(L)}_{\VEC{k}-\VEC{q}},\nonumber\\
K^s_{11,\VEC{k}}&\equiv\frac{kx}{q}
     P^{(L)}_{\VEC{q}}P^{(L)}_{\VEC{k}-\VEC{q}}S^{(2)}_{\VEC{q},\VEC{k}-\VEC{q}},
\end{align}
where again the integration over $q$ variable is implied and $G^{(2)}$ is the standard kernel of velocity divergence in Eulerian PT (e.g. \cite{Bernardeau:2001qr,Vlah:2012ni}).
In figure \ref{fig:5} we show the results for halo-halo $P_{01}$ modelling, and comparison to the N-body simulation results. 
The model  presented in \ref{eq:P01} contains dark matter parts, $P_{01}$ and $P_{\df\tf}$ that were already extensively discussed in \cite{Vlah:2012ni}.
These two contributions we actually measure from the simulations in order to focus on the bias modelling and reduce potential degeneracy with PT modelling.
It is also good to note that on the scales of interest, using $P_{\df \tf}$ either measured from simulations or predicted by PT has only a slight impact on 
best fit value of $b^{01}_2$.
We use the $b_1$ values determined from the matter-halo power spectra $P^{(hm)}_{00}$ from previous section. Note that the relative contributions 
from $b_2$ and $b_{3,nl}$ terms in $P^{(h \bar{h})}_{01,\VEC{k}}$ differ from those in $P^{(h \bar{h})}_{00,\VEC{k}}$, but the scale dependence of these terms 
is similar over the limited range of scale of interest here ($0.05h/Mpc<k<0.2h/Mpc$). This again suggests we can replace all the nonlinear bias terms 
with a single effective $b_2^{01}$, which however can take a different numerical value from $b_2^{00}$. 
We thus fit for a new set of $b_2^{01}$ values in order to achieve better correspondence with the halo simulation results.
In the same figure we also show the results using just the linear biasing $b_1$. 

%=================================%
\subsubsection{On the different $b_2$ values for $P_{00}^{(hm)}$ and $P_{01}^{(hh)}$}
%=================================%

As was mentioned above two different sets of $b_2$ parameters are used: 
one set to model matter halo-cross power spectrum $P^{(hm)}_{00}$ 
and, consequently, halo-halo auto power spectra $P^{(hh)}_{00}$, 
and the second set to model the predictions for $P^{(hh)}_{01}$. 
We have argued that this is necessary because there are several free bias 
parameters that enter the power spectra at 1-loop level, all of which 
have a similar scale dependence, which means we cannot determine them 
individually and we have replaced them with a single effective $b_2$ parameter instead. 
However, while for each statistic we can 
replace them all with a single effective parameter, the relative contributions from each 
physical nonlinear bias to 
different statistics changes, so the values of the effective parameters 
can change as well. 

\begin{figure}[t!]
    \centering
    \includegraphics[scale=0.4]{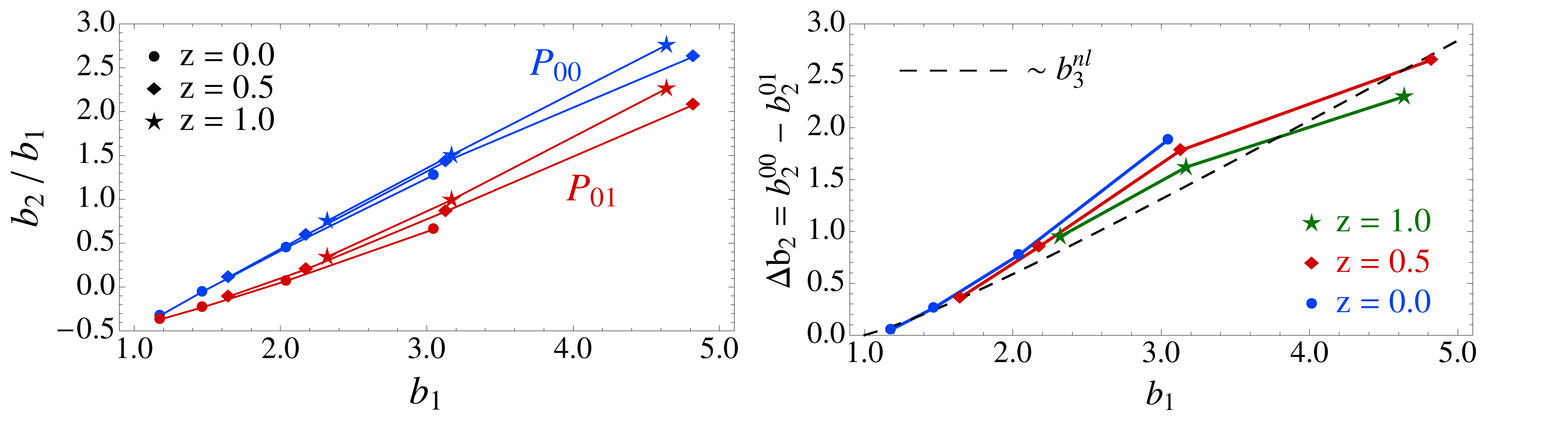}
    \caption{\small  Comparison of $b_2^{00}$ and $b_2^{01}$ used in modelling of $P_{00}$ and $P_{01}$ terms.
    In the left panel we show redshift dependence of  $b_2/b_1$ values for $P_{00}$ term (blue) and $P_{01}$ (red) term.
    In the right panel we show the dependence of the difference $\Delta b_2=b^{00}_2-b^{01}_2$ on the values of $b_1$. We show results for
    redshift $z=0.0$ (blue), $z=0.5$ (red) and $z=1.0$ (green).}
    \label{fig:6}
\end{figure}

One wishes to 
have a biasing model valid for all the statistics of interest and thus for all 
the correlators used in our RSD model. Ideally this would also include the 
higher order correlations, as well as all correlations with the dark matter. 
It is important to realize that this 
difference in $b_2$ bias stays at the level of $P_{00}$ and $P_{01}$ correlators, 
since all the higher order correlators come only through these two terms, at least 
at one loop level we work here. So one can ask whether one can explain the 
difference using a physically motivated model of nonlinear biasing. 
We find that $b_s$ does not seem to matter much assuming its values from the 
bispectrum analysis \cite{Baldauf:2012hs,Chan:2012jj}, or its values from the coevolution model assuming 
it is zero in initial conditions, and so can be ignored. In contrast, 
$b_2$ and $b_{3,nl}$ appear to be equally important \cite{McDonald:2009dh, Saito:2013}.
For the terms we are discussing, this gives
\begin{align}
     P^{(hh)}_{00,\VEC{k}}=&b_1^2 P_{00,\VEC{k}}+2b_1\Big[ b_2
     K_{00,\VEC{k}}+b_sK^s_{00,\VEC{k}}\Big]+\frac{1}{2}\Big[ b^2_2
     K_{01,\VEC{k}}+ b^2_s K^s_{01,\VEC{k}}\Big]\nonumber\\
     &+b_2b_s K^s_{02,\VEC{k}}+2b_{3,nl}\sigma_{3,k}^2P^{(L)}_k,
\end{align}
and also
\begin{align}
    P^{(hh)}_{01,\VEC{k}}=&b_1^2P_{01,\VEC{k}}+b_1(1-b_1)\alpha P_{\df\tf,\VEC{k}}+
    \alpha\Big[b_2K_{10,\VEC{k}}+b_sK^s_{10,\VEC{k}}\Big]\nonumber\\
    &+\alpha b_1\Big[b_2K_{11,\VEC{k}}+b_sK^s_{11,\VEC{k}}\Big]
    +\alpha b_{3,nl}\sigma_{3,k}^2P^{(L)}_k,
\end{align}
where $\sigma_{3,k}$ is defined in \cite{McDonald:2009dh} and for detailed 
discussion of these terms we refer to \cite{Saito:2013}.
The coevolution model predicts a specific value for $b_{3,nl}$ today assuming it is 
zero initially \cite{Chan:2012jj}. Using these predicted values we can predict the difference between 
the two effective $b_2$ values: we find that this model indeed predicts that 
$b_2^{00}$ is larger than $b_2^{01}$ and that the amplitude of the difference increases with 
halo bias $b_1$. The prediction of this model is shown in figure \ref{fig:6}, together with the 
results from the fits. We show the redshift dependence of $b_2$ on $b_1$ and the
difference $\Delta b_2=b^{00}_2-b^{01}_2$.

Even having these three nonlocal bias models may not be all that is required for a 
complete model: in the peak model one expects to have $k^2$ corrections to linear bias both in the 
halo density and in the halo velocity in the initial conditions \cite{Desjacques:2008jj, Desjacques:2010gz}. 
If shown to be significant at later times of evolution, this could 
play an important role in modelling of all velocity correlators. 

%===============================================================%
\subsection{$P^{(hh)}_{02}$ term}
%===============================================================%

Next term we are considering is the correlation of the halo field with the halo kinetic energy density field. 
This term has the only scalar part but these give the contributions to both $\mu^2$ and $\mu^4$ 
parts of the total redshift power spectrum. We have 
\begin{align}
\la \df^{h} | T_2^{\bar{h}} \ra &= b_1\la \df|u^2_\pp\ra+b_1 \bar{b}_1\la \df |\df
u^2_\pp\ra + \frac{1}{2}b_2\la\df^2|u_\pp^2\ra +\frac{1}{2}b_s\la s^2|u_\pp^2\ra\nonumber\\
&+\frac{1}{2}\bar{b}_1b_2\la\df^2|\df u_\pp^2\ra
+\frac{1}{2}b_1\bar{b}_2\la\df|\df^2 u_\pp^2\ra_c
+\frac{1}{4}b_2\bar{b}_2\la\df^2|\df^2 u_\pp^2\ra_c
\end{align}
where we have used the same renormalization scheme as for $P^{(hh)}_{00}$ term. Subscript $c$ 
stands for the connected part of the correlator, while the disconnected parts of this terms get renormalized.
We get for the power spectrum of this term
\begin{align}
P^{(h \bar{h})}_{02,\VEC{k}}=b_1 \bar{P}_{02,\VEC{k}} +
P^{hh}_{00,\VEC{k}}\sigma_v^2-\Big[b_2K_{20,\VEC{k}}+b_sK^s_{20,\VEC{k}}\Big],
\label{eq:P02}
\end{align}
where the first term is the reduced dark matter contribution $\propto\la\df|v_\pp\ra_c$ as defined in \cite{Vlah:2012ni}.
Nonlinear bias comes in through the biasing in $P^{hh}_{00}$ term and we also have the contributions 
of two additional terms 
\begin{align}
K_{20,\VEC{k}}&\equiv\frac{q_\pp}{q^2}\frac{(\VEC{k}-\VEC{q})_\pp}{(\VEC{k}-\VEC{q})^2}
     P^{(L)}_{\VEC{q}}P^{(L)}_{\VEC{k}-\VEC{q}},\nonumber\\
K^s_{20,\VEC{k}}&\equiv\frac{q_\pp}{q^2}\frac{(\VEC{k}-\VEC{q})_\pp}{(\VEC{k}-\VEC{q})^2}
     P^{(L)}_{\VEC{q}}P^{(L)}_{\VEC{k}-\VEC{q}}S^{(2)}_{\VEC{q},\VEC{k}-\VEC{q}}.
\end{align}
These terms have contributions to both $\mu^2$ and $\mu^4$. As was shown in \cite{Vlah:2012ni},
the term that is proportional to the velocity dispersion will appear in the $\mu^2$ part. 
Since for DM it is not possible to evaluate the total velocity dispersion using PT
we had to model small scale contributions adding a part which was motivated by the velocity dispersion in a halo using the
halo model. For halos the situation is simpler in that we do not expect such a term to be present, as  
there is no small scale velocity dispersion contribution, and hence no FoG. Most of the contribution
comes from the large scale velocities  which can at least in principle be described using PT. 

%=================================%
\subsubsection{Velocity dispersion predictions and simulation measurements}
%=================================%

\begin{figure}[t!]
    \centering
    \includegraphics[scale=0.52]{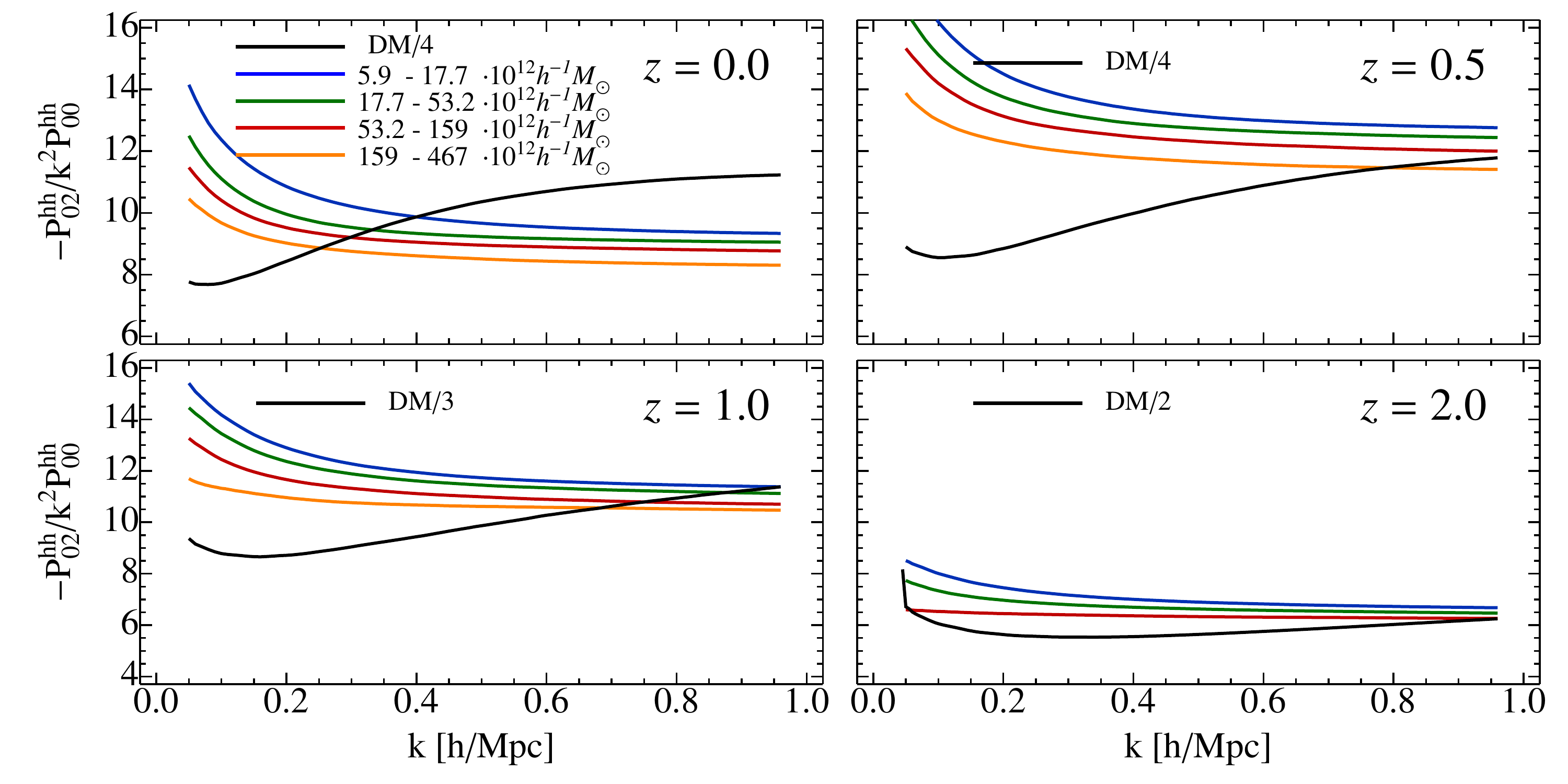}
    \caption{\small $P_{02}/P_{00}$ for halos and dark matter.
    Results are presented for dark matter halos (color coding for respective bias values) and for the dark matter (in black). 
    Values of velocity dispersion can be estimated at the smaller scale regions (higher $k$), where the ratio tends approach the constant values.
    We show the ratio for four different halo mass bins and for four different redshifts.}
    \label{fig:7}
\end{figure}

\begin{figure}[t]
    \centering
    \includegraphics[scale=0.55]{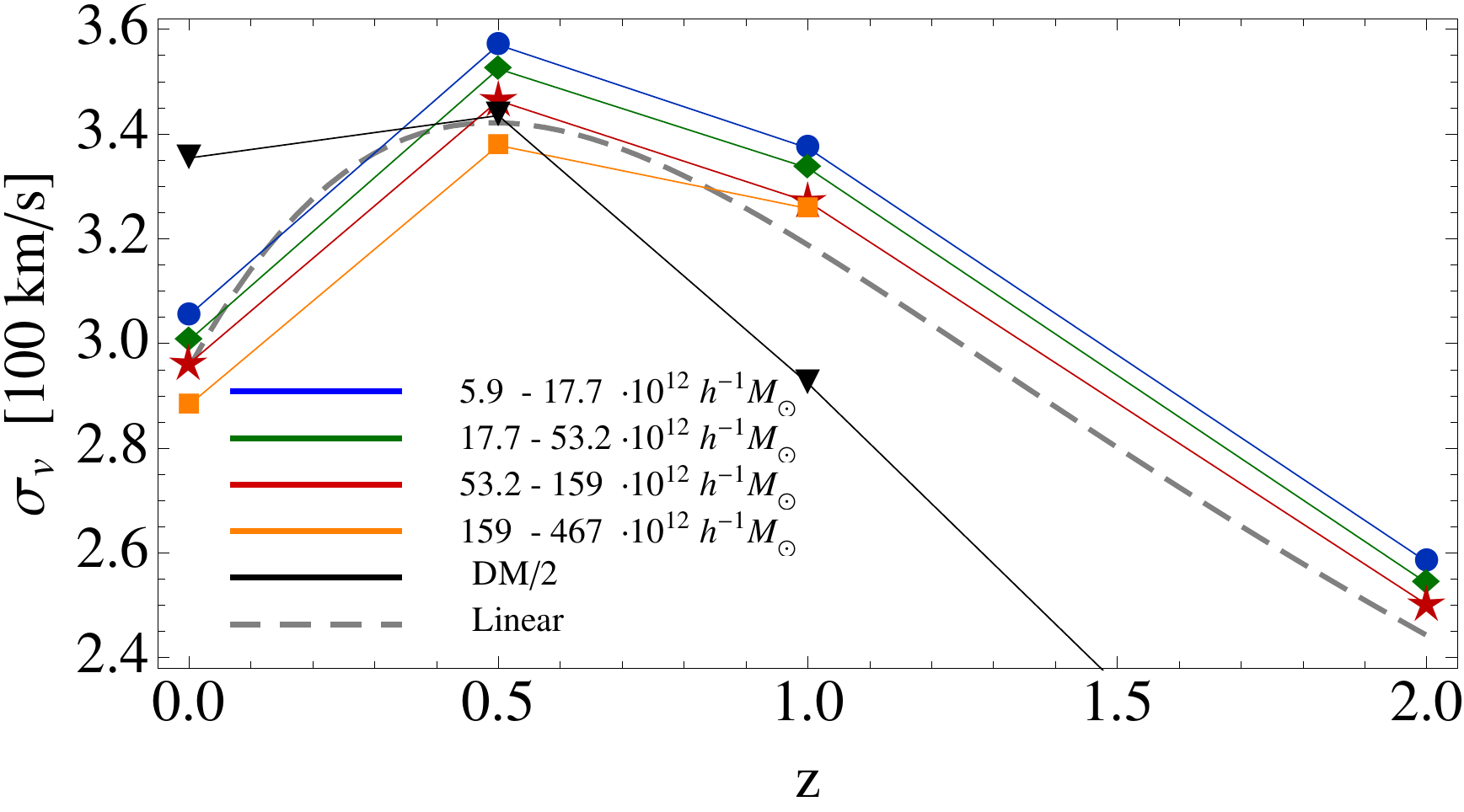}
    \caption{\small Redshift dependence of the halo velocity dispersion measurements obtained from simulations.
    Four different mass bins values are shown (in different color coding respectively). We also show the velocity dispersion measurements for dark matter case (in Black).
    Also shown are the linear theory prediction for velocity dispersion values (gray dashed line).
   Notice that the values of measured velocity dispersion have a weak dependence on mass of the halo bins.}
    \label{fig:8}
\end{figure}

In order to test the velocity dispersion behaviour in more detail let us consider more closely the contribution
of $\mu^2$ part of $P^{hh}_{02}$. From previous work on dark matter \cite{Vlah:2012ni},
we have seen that on quasi-nonlinear scales most of the contributions to the $P_{02}$ term came from 
the part proportional to the factor corresponding to the dark matter velocity dispersion multiplied with $P_{00}$ term. 
For halos this will be generalized to the $\sigma^2_v P^{hh}_{00}$, where we have the halo velocity dispersion instead. 
In order to test the behaviour on the halo velocity dispersion in simulations, in figure \ref{fig:7} we show the $k$ dependence of
$P^{hh}_{02}/P^{hh}_{00}$. From the figure it can be seen that at smaller scales this ratio tends to a constant 
for all mass bins and redshifts. This constant values can be interpreted as the halo velocity dispersion $\sigma^2_v$. 
Note that for this figure we include the shot noise $\bar{n}^{-1}$ in $P^{hh}_{00}$, and we expect 
there is also the shot noise $\sigma_v^2/\bar{n}$ in $P_{02}$.

In figure \ref{fig:8} we show the redshift dependence of the halo velocity dispersion $\sigma_v$ as obtained from simulations (high-k regime in figure \ref{fig:7}).
We see the trend of the velocity dispersion values increasing and then decreasing with redshift, 
as predicted by the linear theory prediction of the velocity dispersion
\begin{equation}
\sigma_{v,\text{lin}}^2(\tau)=\frac{1}{3}\int{\frac{d^3q}{(2\pi)^3}\frac{P_{\tf\tf}(q,\tau)}{q^2}}.
\end{equation}
We see that velocity dispersions weakly depend on the masses of the halos, 
at all the redshifts, and are in a good agreement with linear theory predictions. 
In contrast, the dark matter velocity dispersion is significantly higher than the linear theory prediction. 
Slight dependence of halo velocity dispersion on halo mass, noticeable in figures \ref{fig:7} and \ref{fig:8} 
can be explained by the terms in equation \ref{eq:P02} that do not depend on $\sigma_v$. Higher order terms 
can also play a role in modelling both $\left\langle \df^h | u_\pp^2\right\rangle$ and also 
connected term $\left\langle \df^h | \df^h u_\pp^2\right\rangle$. 

\begin{figure}[t!]
    \centering
    \includegraphics[scale=0.29]{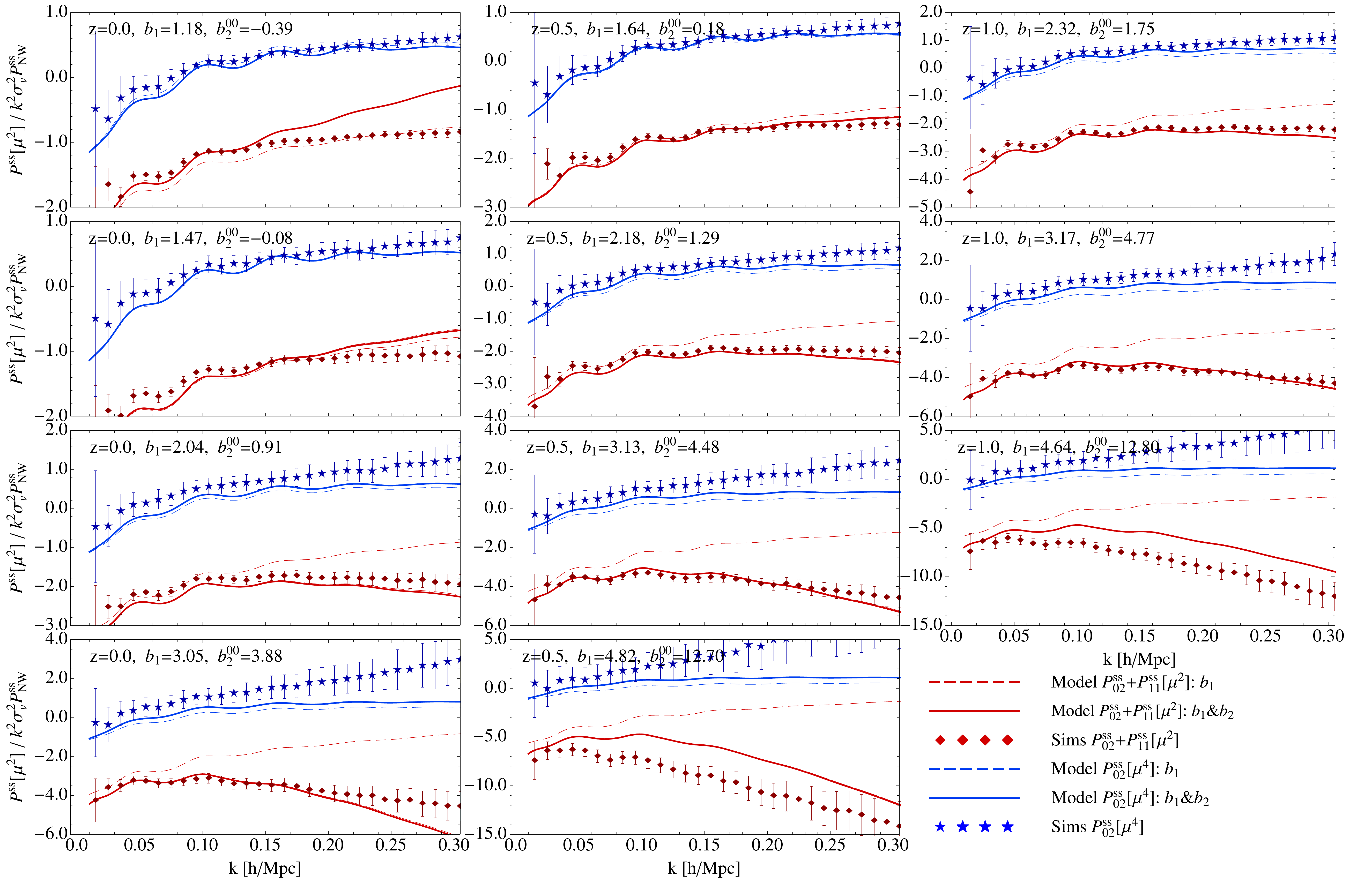}
    \caption{\small $\mu^2$ part of $P_{02}+P_{11}$ (red color) and $\mu^4$ part of $P_{02}$ (blue color) is shown at three different redsifts,
    for several mass bins. Simulation measurements (points) of these two terms are compared to the model \ref{eq:P02} using linear (dashed lines) and 
    non-linear (solid lines) biasing.  All the plots are divided by the linear predictions $k^2\sigma_v^2P_{NW}$ with no BAO wiggles.}
    \label{fig:9}
\end{figure}

In figure \ref{fig:9} we show the $\mu^2$ part of $P^{(hh)}_{02}+P^{(hh)}_{11}$ and $\mu^4$ part of $P^{(hh)}_{02}$ as contributions to the 
total redshift power spectrum. The sum of $P^{(hh)}_{02}$ and $P^{(hh)}_{11}$ terms is chosen since both of these terms have isotropic 
shot noise contributions $\sigma^2/\bar{n}$, and which can be large for rare tracers such as halos. 
Since these two contributions come with opposite signs and exactly cancel in the sum, they give no residual shot noise contribution to the 
total RSD power spectrum . From equation \ref{eq:P02}  we see that $P^{{hh}}_{02}$ term also contains contributions 
from $P^{{hh}}_{00}$ term which contains nonlinear $b_2$ bias. Thus for modelling the $P^{{hh}}_{02}$ term we use the same $b_2$ values as 
for $P^{{hh}}_{00}$ term. 

%===============================================================%
\subsection{$P^{(hh)}_{11}$ term}
%===============================================================%

Next term to consider is the autocorrelation of the halo momentum field $P^{(hh)}_{11}$. This term will give the contributions to
both $\mu^2$ (which has only contributions from the vector part) and $\mu^4$ angular dependence (which has dominantly contributions from the scalar part
on large scales, as it is present already in linear theory, while the vector part becomes important on small scales).
Leading term on the large scales has $\mu^4$ angular dependence and 
in the limit of small $k$ consists only of the third Kaiser term. 
Correlating the two halo momentum fields using the biasing model (equation \ref{eq:halofield2}) we get
\begin{align}
\la T_1^{h} | T_1^{\bar{h}} \ra &= \la u_\pp|u_\pp\ra+(b_1+\bar{b}_1)\la u_\pp |\df u_\pp\ra + b_1\bar{b}_1\la \df u_\pp |\df u_\pp\ra \nonumber\\
&+\frac{1}{2}(b_2+\bar{b}_2)\la u_\pp| \df^2 u_\pp \ra_c + \left\lbrace b_1,b_2\right\rbrace \la\df u_\pp|\df^2 u_\pp\ra_c+
\frac{1}{4}b_2 \bar{b}_2 \la\df^2 u_\pp|\df^2 u_\pp\ra_c
\label{eq:P11v1}
\end{align}
where we can again use the same renormalization scheme as for $P^{(hh)}_{00}$ term. Subscript $c$ again represents
connected part of the correlator, while the disconnected parts gets renormalized.
Keeping the terms at one loop order we get for the power spectrum, 
\begin{align}
P^{(h_1h_2)}_{11,\VEC{k}}=P_{11,\VEC{k}}+\big[(b_1-1)+(\bar{b}_1-1)\big]\frac{\mu}{k}B_{11,\VEC{k}}+(b_1\bar{b}_1-1)C_{11,\VEC{k}},
\label{eq:P11v2}
\end{align}
where the dark matter terms $P_{11}$, $B_{11}$ and $C_{11}$ used here have been computed in \cite{Vlah:2012ni}.
At the one loop level of PT modelling, there are no contributions of $b_2$ bias terms, 
as can be seen from \ref{eq:P11v1} where all $b_2$ terms appear only at the 2-loop level. The same is true for 
higher order local and nonlocal bias contributions.

\begin{figure}[t!]
    \centering
    \includegraphics[scale=0.3]{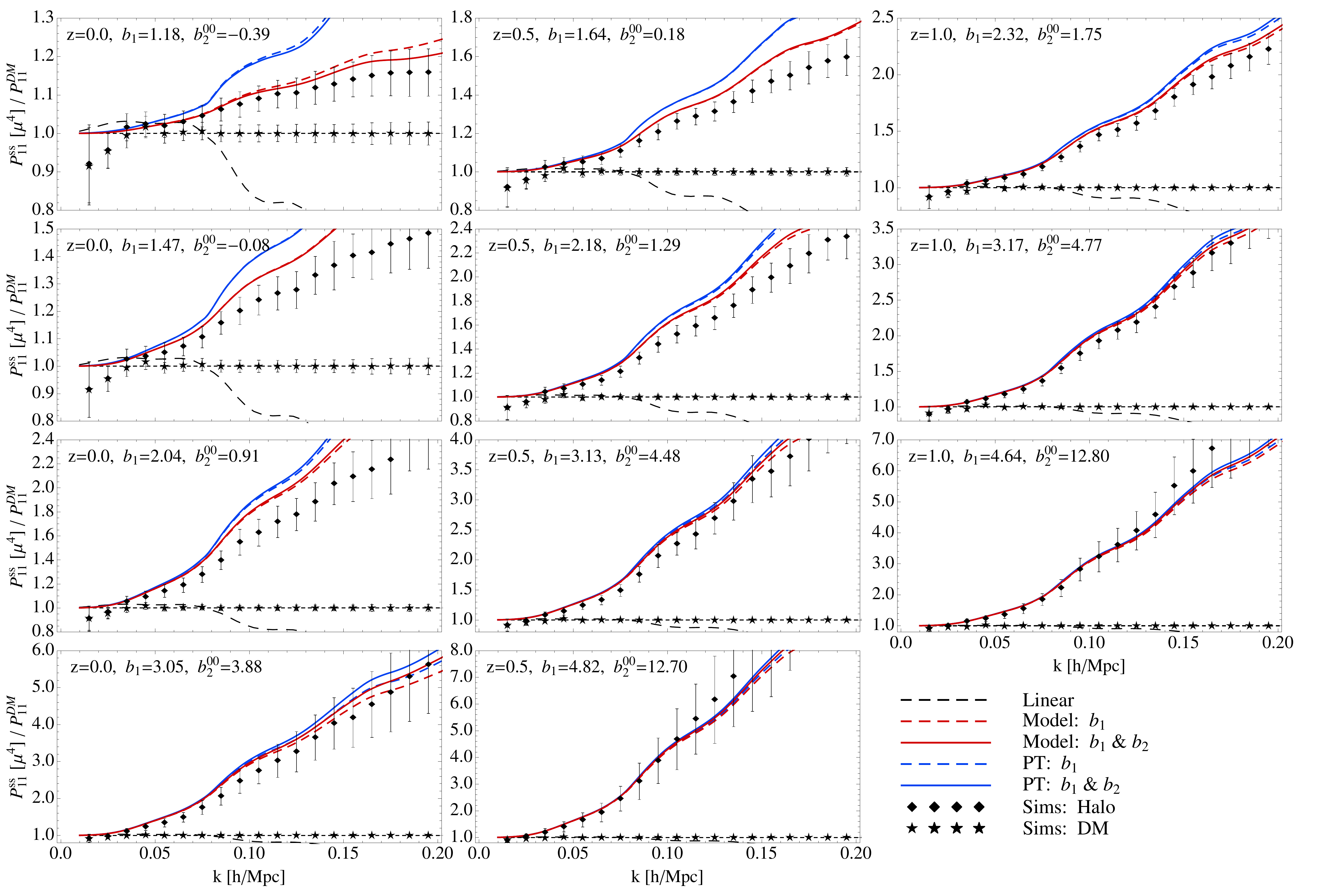}
    \caption{\small $\mu^4$ part of the halo momentum power spectrum $P^{(hh)}_{11}$ relative to $P_{11}$ for dark matter. Results are shown for halos for several mass bins at 
    redshifts $z=0.0,~0.5,~1.0$. We show the full SPT result (blue line) and the model (red line) where simulations are used for the DM part of $P_{11}$.
    Both the linear biasing model (dashed lines) and the nonlinear biasing 
   model with $b_2$ terms is shown (solid line). Here we emphasize that the $b_2$ parameters 
    are not free, but have been fixed by the $P_{00}$ and $P_{01}$ analyses. We also show the halo simulation 
    measurements (black dots) and $b_1$ times DM simulations (black stars). 
    Kaiser linear result with $b_1$ bias is also shown (lond-dashed black line).}
    \label{fig:10}
\end{figure}

In figure \ref{fig:10} we show the scalar part of velocity moment halo power spectra $P^{(hh)}_{11}$. To improve the precision
we use simulations for the scalar part of  $P_{11}$ for dark matter. 
For a more extensive discussion about how to model this in PT and consistency of this procedure we refer to \cite{Vlah:2012ni}.
We also investigate the two loop contributions and show terms proportional to $b_2$ in equation \ref{eq:P11v1}.
Notice that these contributions are suppressed by the $f^2D^6$ factor so the contribution at higher redshifts quickly becomes less important.
Overall we see that the model qualitatively reproduces simulations well, but not quantitatively for all halos. 
Notice the very strong scale dependence of the term, specially for the highly biased halos. The scale dependence is induced already at the 
linear biasing level. 
As mentioned in previous section, vector part of $P^{(hh)}_{11}$ is shown in figure \ref{fig:9} where it is combined with $\mu^2$ part of the $P^{(hh)}_{02}$
term. Overall we conclude that our model reproduces all the trends seen in simulations without any additional free parameters (except the $b_1$ and $b_2$ bias 
parameters that have been fitted to measurement of $P_{00}^{(hh)}$ and $P_{01}^{(hh)}$), 
but is not sufficient to achieve high precision predictions down to very small scales. 

%===============================================================%
\subsection{$P^{(hh)}_{12}$ and $P^{(hh)}_{03}$ terms}
\label{subsec:1203}
%===============================================================%

\begin{figure}[t!]
    \centering
    \includegraphics[scale=0.3]{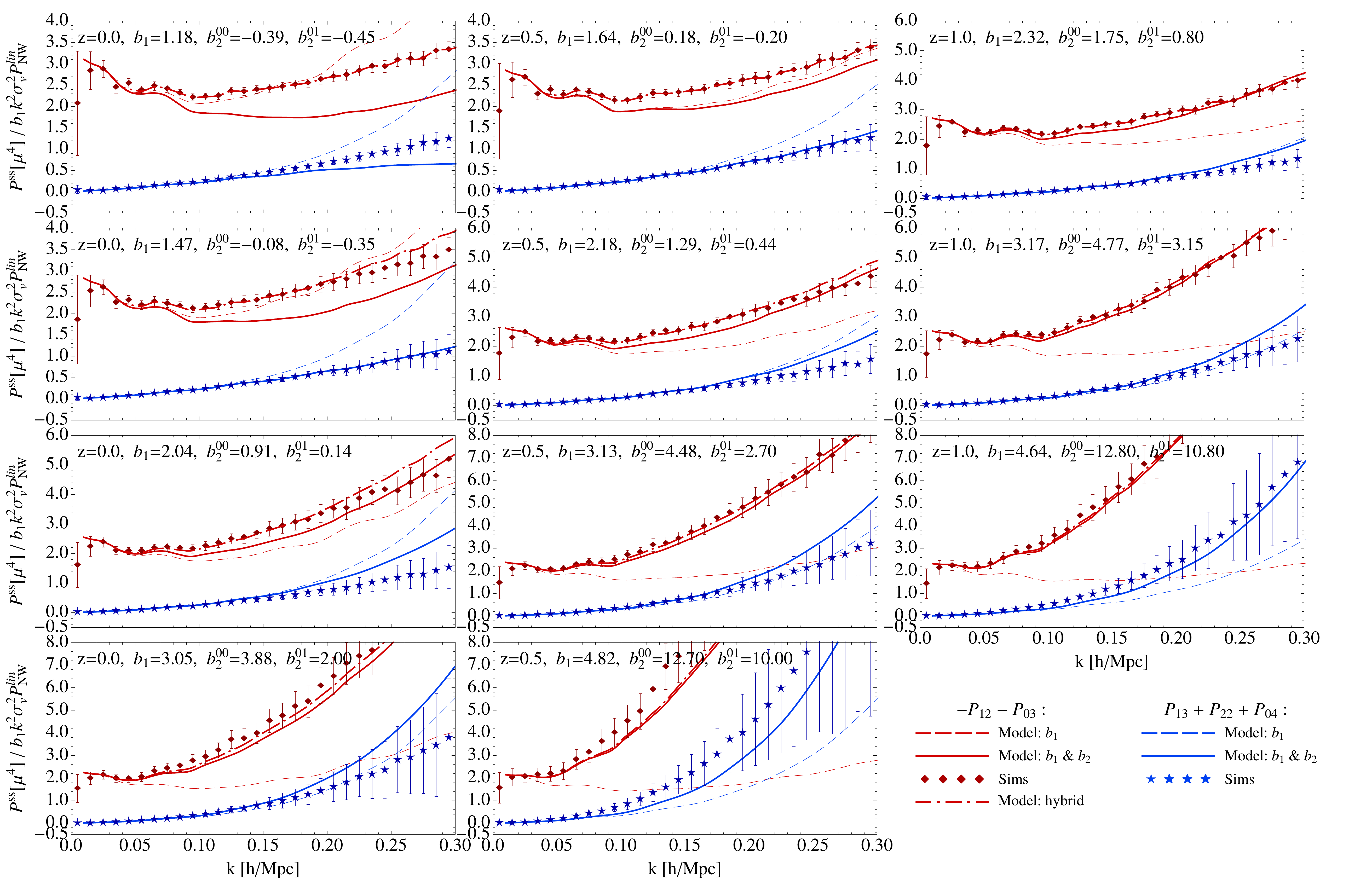}
    \caption{\small $\mu^4$ parts of the $P_{12}+P_{03}$ (red) term 
    and $P_{13}+P_{22}+P_{04}$ (blue) for several mass bins at redshifts $z=0.0,~0.5,~1.0$. 
    We take the specific combinations for which the shot noise contributions cancel out. 
    We show results for the linear biasing model (dashed) and for the nonlinear biasing  (solid). Note the large effect of
    nonlinear biasing. Here again we emphasize that the $b_2$ parameters are not free, but have been fixed by the $P_{00}$ and $P_{01}$ analyses. 
    For $P_{12}+P_{03}$ term we also present a hybrid result (red dot-dashed line) where we fit for the correction in the lowest mass bin at each redshift 
    and apply it to higher mass bins. All the plots are divided by the linear predictions $b_1 k^2\sigma_v^2P_{NW}$ with no BAO wiggles.}
    \label{fig:10.5}
\end{figure}

In this subsection we look at higher order moment terms $P^{(hh)}_{12}$ and $P^{(hh)}_{03}$ which have the lowest angular contribution at $\mu^4$ level. We will only model  the lowest order $\mu^4$ terms, which give the nonlinear corrections to the last Kaiser term. Nonlinear $b_2$ bias enter in these terms indirectly through the
terms $P^{(hh)}_{00}$, $P^{(hh)}_{01}$ and consequently $P^{(hh)}_{11}$. 
For the correlation of first and second order momentum fields $P^{(h \bar{h})}_{12}$ we have
\begin{align}
\la T_1^{h} | T_2^{\bar{h}} \ra &= \la (1+\df^h ) u_\pp|(1+\df^{\bar{h}})u_\pp^2\ra \nonumber\\
&= \la (1+\df^h ) u_\pp| u_\pp^2\ra + \la (1+\df^h) u_\pp|\df^{\bar{h}} u_\pp^2\ra.
\end{align}
By decomposing these two terms further we get
\begin{align}
\la (1+\df^h ) u_\pp| u_\pp^2\ra &= \la (1+\df ) u_\pp| u_\pp^2\ra + \la (\df^h-\df) u_\pp| u_\pp^2\ra \nonumber\\
%&=  \la (1+\df ) u_\pp| (1+\df )u_\pp^2\ra + \la (\df^h-\df) u_\pp| u_\pp^2\ra -\la (1+\df ) u_\pp| \df  u_\pp^2\ra\nonumber\\
&=  \la (1+\df ) u_\pp| (1+\df )u_\pp^2\ra + (b_1-1)\la \df u_\pp| u_\pp^2\ra -\la (1+\df ) u_\pp| \df  \ra  \la u_\pp^2 \ra+\ldots\nonumber\\
&=(2\pi)^3\left( P_{12,\VEC{k}} -i (b_1-1)B_{12,\VEC{k}}+P_{01,\VEC{k}}\sigma_v^2 +\ldots \right)\df^D(\VEC{k}-\VEC{k}'),\nonumber\\
\la (1+\df^h) u_\pp|\df^{\bar{h}} u_\pp^2\ra & =\la (1+\df^h) u_\pp|\df^{\bar{h}} u_\pp^2\ra_c + \la (1+\df^h) u_\pp|\df^{\bar{h}} \ra\la u_\pp^2\ra\nonumber\\
&=(2\pi)^3\left( -P^{(h\bar{h})}_{01,\VEC{k}}\sigma_v^2\right)\df^D(\VEC{k}-\VEC{k}'),
\end{align}
where we again refer to \cite{Vlah:2012ni} for detailed expressions for dark matter terms $P_{12}$  and $B_{12}$, and we use the property of high order momentum
correlators $P_{LL'}=P^*_{L'L}$. Connected part will have contributions at the level higher than one loop so will not be considered here. 
Combining these results above we get 
\begin{align}
P^{(h \bar{h})}_{12,\VEC{k}}=P_{12,\VEC{k}}-i(b_1-1)B_{12,\VEC{k}}-\left(P^{(h\bar{h})}_{01,\VEC{k}}-P_{01,\VEC{k}}\right)\sigma_v^2.
\end{align}
Similarly for the $P^{(h \bar{h})}_{03}$ term we have,
\begin{align}
\la \df^{h} | T_3^{\bar{h}} \ra &=\la \df^h | (1+\df^{\bar{h}}) u_\pp^3\ra\nonumber\\
&= b_1\la \df |u_\pp^3\ra + b_1 \bar{b}_1 \la \df | \df u_\pp^3\ra+\ldots,
\end{align}
which can be generalized to the following form 
\begin{align}
P^{(h \bar{h})}_{03,\VEC{k}}=3P^{(h\bar{h})}_{01,\VEC{k}}\sigma_v^2,
\end{align}
where we have again omitted the connected part since it gives only contributions at the higher level than one loop.

It is again convenient to combine some of these terms, such as $P_{12}$ and $P_{03}$ to eliminate the shot noise contributions. 
These two terms together give the contribution to the total redshift power spectrum
\begin{align}
 \frac{i}{3}\left(\frac{k \mu}{\mathcal{H}}\right)^3P^{(h\bar{h})}_{03,\VEC{k}}-i\left(\frac{k \mu}{\mathcal{H}}\right)^3P^{(h\bar{h})}_{12,\VEC{k}}=
 i\left(\frac{k \mu}{\mathcal{H}}\right)^3\left(2\sigma_v^2P^{(hh)}_{01,\VEC{k}}+ib_1B_{12,\VEC{k}}-(P_{12,\VEC{k}}+iB_{12,\VEC{k}}+\sigma_v^2 P_{01,\VEC{k}})\right).
\label{eq:P0312}
\end{align}
In figure \ref{fig:10.5} we show the result of this model for sum of two terms. We show contribution from both linear biasing and nonlinear $b_2$
contributions that come through the $P^{(hh)}_{01}$ term and compare it to the simulation measurements. 
We find that nonlinear biasing terms in most cases improve the agreement. Note that the $b_2$ terms are not fitted, but have been fixed by the lower order analyses. 
We also present the result of so called hybrid model where we first fit for the correction to our PT model relative to the simulations in the lowest mass bins at each redshift. 
This correction is applied then to the higher mass bins at the given redshift. We see that this procedure gives an improvement for all mass bins we consider. This suggests 
that improving the modelling of dark matter rather then biasing might be more important in order to improve the result for $P_{12}+P_{03}$.

%===============================================================%
\subsection{$P^{(hh)}_{13}$, $P^{(hh)}_{22}$ and $P^{(hh)}_{04}$ terms}
%===============================================================%

The remaining terms to consider at the $\mu^4$ level to the total RSD power spectrum 
are $P^{(hh)}_{13}$, $P^{(hh)}_{22}$ and $P^{(hh)}_{04}$. First we look at the
$P^{(h \bar{h})}_{13}$ term, 
\begin{align}
\la T_1^{h} | T_3^{\bar{h}} \ra &= \la (1+\df^{h} )u_\pp | (1+\df^{\bar{h}} ) u_\pp^3\ra \nonumber\\
&=\la u_\pp |u_\pp^3\ra + b_1 \la \df u_\pp | u_\pp^3\ra +\bar{b}_1 \la u_\pp | \df u_\pp^3\ra + b_1 \bar{b}_1 \la \df u_\pp | \df u_\pp^3\ra+\ldots,
\end{align}
which can be collected to give
\begin{align}
P^{(h \bar{h})}_{13,\VEC{k}}=&3\sigma_v^2\Big[P_{11,\VEC{k}}-\big[(b_1-1)+(\bar{b}_1-1)\big]\frac{\mu}{k}B_{11,\VEC{k}}+(b_1\bar{b}_1-1)C_{11,\VEC{k}}+\ldots\Big]\nonumber\\
                            =&3P^{(h\bar{h})}_{11,\VEC{k}}\sigma_v^2.
\end{align}
The obtained result is given in terms of previous $P^{hh}_{11}$ term, and velocity dispersion.

Next we look at the  $P^{(h \bar{h})}_{22}$ term,
\begin{align}
\la T_2^{h} | T_2^{\bar{h}} \ra &=\la u_\pp^2|(1+\df^{\bar{h}} )u_\pp^2\ra+\la \df^{h} u_\pp^2|(1+\df^{\bar{h}} )u_\pp^2\ra\nonumber\\
&= \la u_\pp^2 |u_\pp^2\ra + (b_1+\bar{b}_1) \la u_\pp^2 | \df u_\pp^2\ra + b_1\bar{b}_1 \la \df u^2_\pp| \df u^2_\pp \ra+\ldots
\end{align}
which can be collected to give, 
\begin{align}
P^{(h \bar{h})}_{22,\VEC{k}}&=\bar{P}_{22,\VEC{k}} +(b_1+\bar{b}_1)\bar{P}_{02,\VEC{k}}\sigma_v^2+P^{(h\bar{h})}_{00,\VEC{k}}\sigma_v^4+
\big(P^{(h\bar{h})}_{00}\circ\bar{P}_{22}\big)_{\VEC{k}}+\ldots\nonumber\\
&=\bar{P}_{22,\VEC{k}} +b_1\bar{P}_{02,\VEC{k}}\sigma_v^2+P^{(h\bar{h})}_{02,\VEC{k}}\sigma_v^2+\big(P^{(h\bar{h})}_{00}\circ\bar{P}_{22}\big)_{\VEC{k}}.
\end{align}
Here we again refer to \cite{Vlah:2012ni} where  the dark matter terms $\bar{P}_{22}$ and $\bar{P}_{02}$ have been computed using the PT.

Lastly we turn to the $P^{(h \bar{h})}_{04}$ term which is formally of the two loop order but it turns up to be significant contribution to the $\mu^4$ 
part of the total redshift space power spectrum,
\begin{align}
\la \df^{h} | T_4^{\bar{h}} \ra &= b_1\la \df |u_\pp^4\ra + b_1 \bar{b}_1 \la \df | \df u_\pp^4\ra+\ldots.
\end{align}
which in terms of power spectrum gives
\begin{align}
P^{(h \bar{h})}_{04,\VEC{k}}=&6b_1 \bar{P}_{02,\VEC{k}}\sigma_v^2+ b_1\bar{b}_1P^{(h\bar{h})}_{00,\VEC{k}}\Big(3\sigma_v^4+\la u_\pp^4\ra_c \Big).
\end{align}
In figure \ref{fig:10.5} we show the result of modelling the sum of these three terms. We show contribution from both linear biasing and nonlinear $b_2$
contributions that come through the $P^{(hh)}_{00}$, $P^{(hh)}_{01}$ and $P^{(hh)}_{11}$ terms. For comparison we show simulation measurements
of these terms. Once again nonlinear biasing dramatically improves the accuracy of the model. 

%===============================================================%
%===============================================================%
\section{Putting it all together: angular dependence and multipole moments}
\label{sec:ADRMM}
%===============================================================%
%===============================================================%

\begin{figure}[t!]
    \centering
    \includegraphics[scale=0.3]{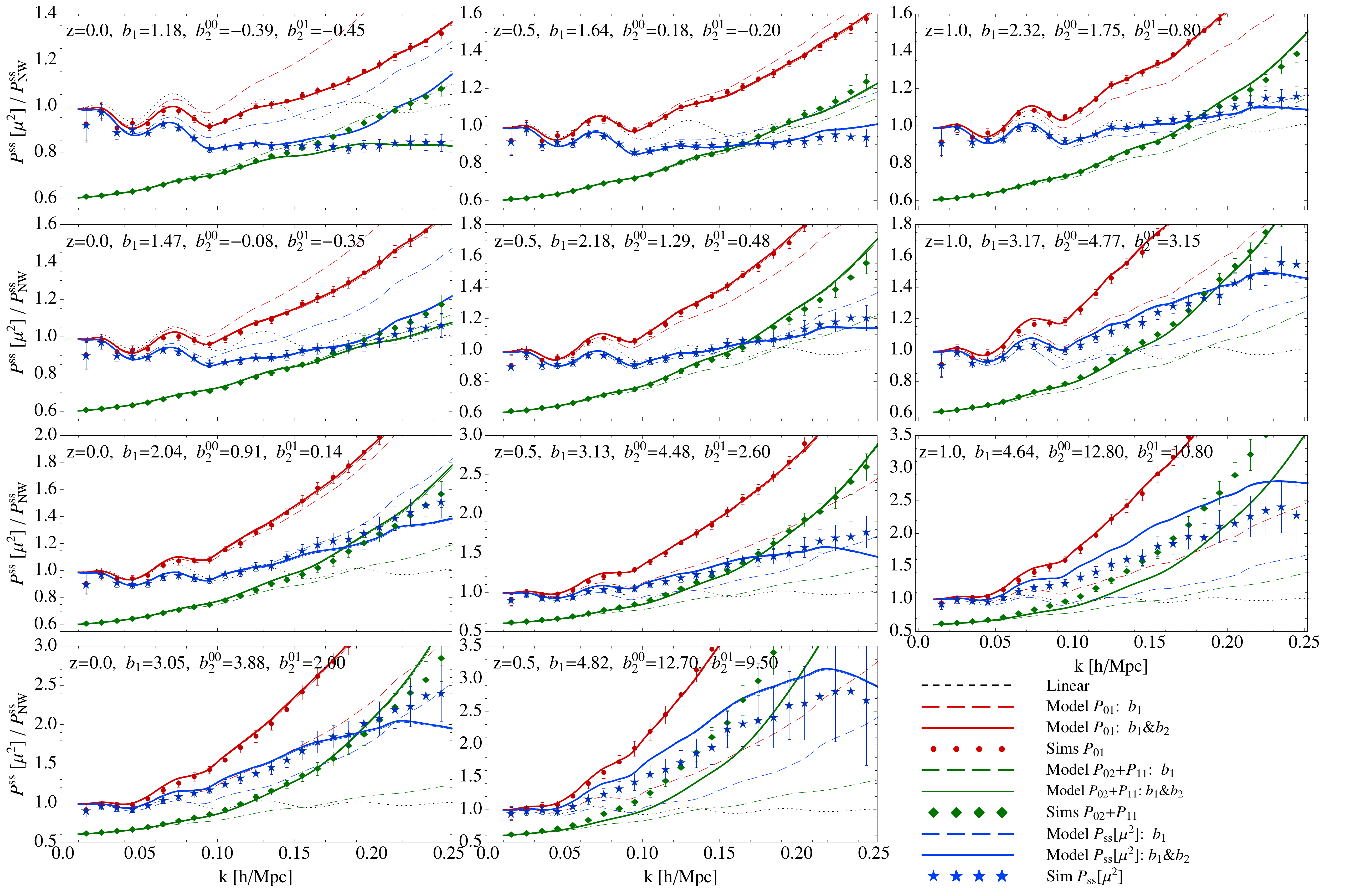}
    \caption{\small Contribution of all the terms to $\mu^2$ part of redshift space power spectrum, for several mass bins at redshifts $z=0.0$,
     $z=0.5$ and $z=1.0$. We also show the contribution of $P_{01}$ term (red) and $P_{02}+P_{11}$ term (green) to the total model (blue).
     $P_{02}+P_{11}$ term has been shifted up by 0.6.
     We compare the model results (solid lines) to the simulation measurements (points). Model where only linear $b_1$ is used is also shown (dashed lines) for comparison. 
     All the lines are divided by no-wiggle Kaiser $\mu^2$ term.}
    \label{fig:11}
\end{figure}

\begin{figure}[t!]
    \centering
    \includegraphics[scale=0.3]{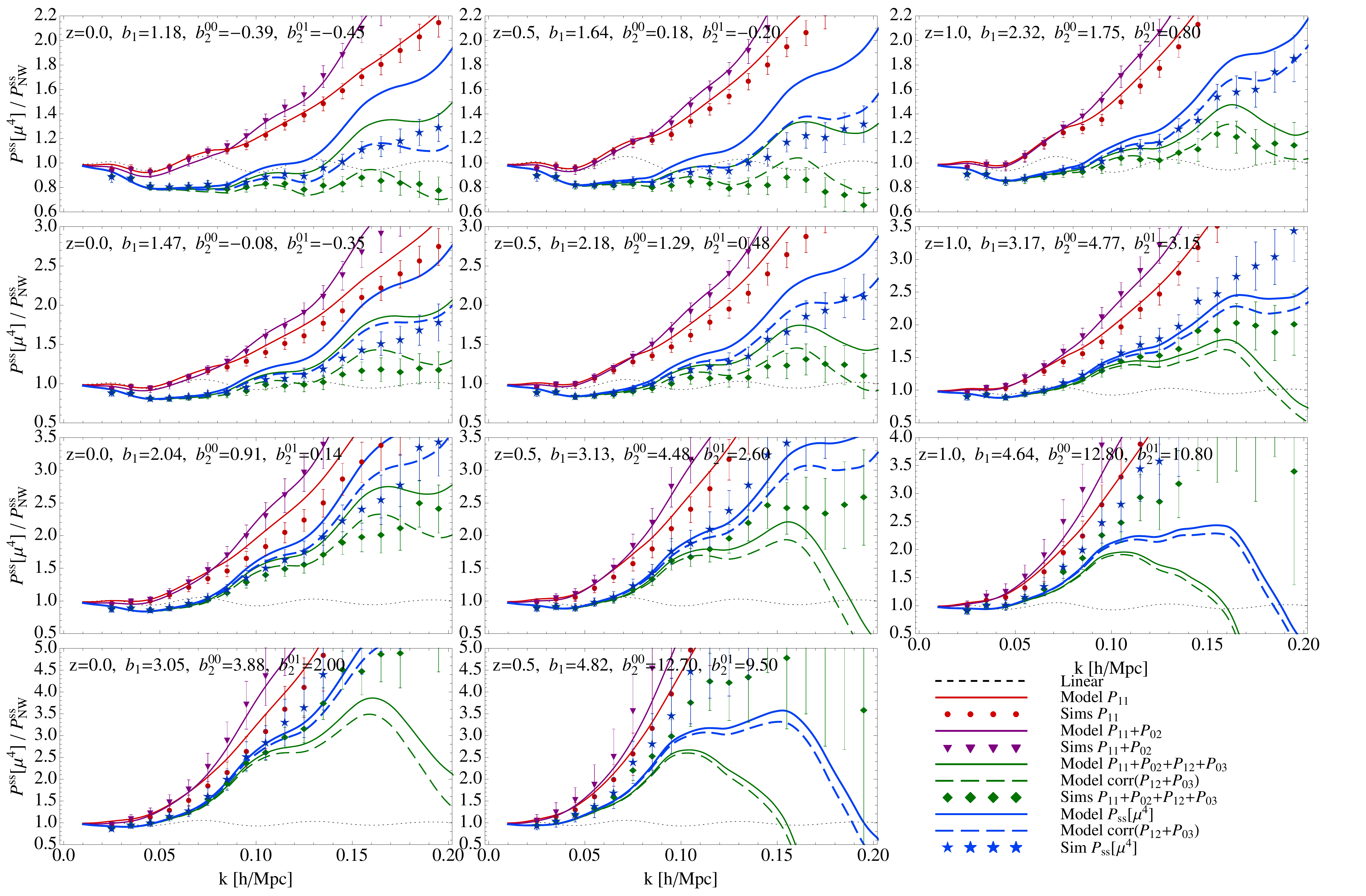}
    \caption{\small Contribution of all the terms to $\mu^4$ part of redshift space power spectrum, for several mass bins at redshifts $z=0.0$,
     $z=0.5$ and $z=1.0$. We also show the contribution of $P_{11}$ term (red), $P_{02}$ term (purple), $P_{12}+P_{03}$ term (green) to the total model (blue).
     We compare the model results (solid lines) to the simulation measurements (points). We also show the results when the correction to $P_{12}+P_{03}$ term
     is added to the model (dashed lines), as discussed in section \ref{subsec:1203}.
     All the lines are divided by no-wiggle Kaiser $\mu^4$ terms.}
    \label{fig:12}
\end{figure}

\begin{figure}[t!]
    \centering
    \includegraphics[scale=0.272]{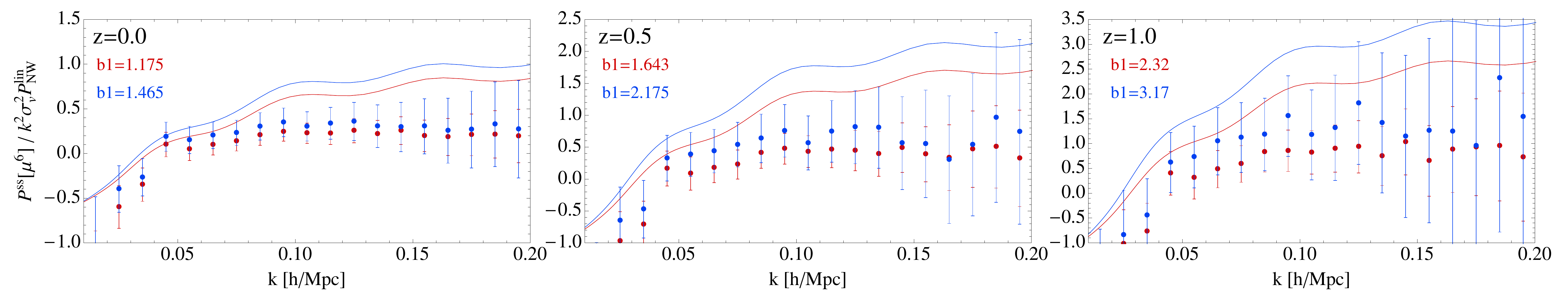}
    \caption{\small Contribution to $\mu^6$ part of redshift space power spectrum, for two mass bins at redshifts $z=0.0$,
     $z=0.5$ and $z=1.0$. We show one loop PT result  (solid lines) and compare to the simulation measurements (points). 
      All the plots are divided by the linear predictions $k^2\sigma_v^2P_{NW}$ with no BAO wiggles.}
    \label{fig:12.5}
\end{figure}

In this section we collect all the contributions to the $\mu^2$ and $\mu^4$ angular dependence in the redshift space power spectrum.
First we collect all of the terms that contribute to the halo redshift space power spectrum up to $\mu^4$
angular dependence.  We write the power spectrum in the form of powers of $\mu^2$, 
\begin{align}
P_{ss,\VEC{k}}=A_k+B_k\mu^2+C_k\mu^4+D_k\mu^6+\ldots, 
\label{eq:nosum}
\end{align}
where the isotropic factor terms are
\begin{align}
A_k&=P_{00,k}\left[\mu^0\right],\nonumber\\
B_k&=P_{01,k}^{ss}\left[\mu^2\right]+P_{02,k}^{ss}\left[\mu^2\right]+P_{11,k}^{ss}\left[\mu^2\right],\nonumber\\
C_k&= P_{11,k}^{ss}\left[\mu^4\right]+P_{02,k}^{ss}\left[\mu^4\right]+P_{12,k}^{ss}\left[\mu^4\right]+P_{03,k}^{ss}\left[\mu^4\right]+P_{13,k}^{ss}\left[\mu^4\right]
+P_{22,k}^{ss}\left[\mu^4\right]+P_{04,k}^{ss}\left[\mu^4\right],\nonumber\\
D_k&= P_{12,k}^{ss}\left[\mu^6\right]+P_{13,k}^{ss}\left[\mu^6\right]+P_{22,k}^{ss}\left[\mu^6\right]+\ldots.
\end{align}
Each of these terms has been modelled separately in previous chapters. For $\mu^6$ part we write down only the terms that contribute at one loop level.
In figure \ref{fig:11} we show the performance of the model on $\mu^2$, and in figure \ref{fig:12} on $\mu^4$. We see that 
the model performs reasonably well on $\mu^2$ term up to $k\approx 0.15$ for most of the halo masses and redshifts. 
To achieve this we had to use the dark matter simulations for the dominant terms ($P_{00}$, $P_{01}$, $P_{11}$ and 
$P_{\delta \theta}$), include nonlinear biasing with 2 nonlinear bias parameters, and take into account the effect of  
halo exclusion in the the stochasticity parameter $\Lambda$. 
Main remaining source of discrepancy here is the $P_{02}+P_{11}$ contribution, so improving this term would lead to a further overall
improvement of the $\mu^2$ part.  On the other hand, our model is less successful for $\mu^4$ part since these terms 
show stronger scale dependence, 
for which our adopted biasing model is less successful. Also, note that $\mu^4$ term has seven constituent terms and since the total error 
on the final result is cumulative, the final discrepancy from simulations tends to be larger than that for $\mu^2$.
Nevertheless, it appears that the main culprit is our modeling of $P_{03}+P_{12}$, which is the analog of $P_{02}+P_{11}$ as it has 
very similar correlators. 
In figure \ref{fig:12.5} we also show the leading, one loop contributions to $\mu^6$. Terms that contribute to $\mu^6$ at one loop level 
are $P_{12}$, $P_{03}$, $P_{13}$ and $P_{22}$. From the figure we see that we can give some quantitative prediction of simulation results,
but higher order modelling is required in order to archive better agreement.  Also note that the prediction that we show is strictly one loop SPT 
with no additional parameters. 

In \cite{Vlah:2012ni} we resummed our model \ref{eq:nosum} to allow a continuation to higher powers of $\mu^2$. 
In the dark matter case this procedure was not possible because of the small scale velocity dispersion, and 
because these small scale velocity dispersions appear with different amplitudes in contributing terms. This is because dark matter 
is distributed into halos of widely varying mass and different terms pick up different mass weightings: for example, some are additionally 
weighted by bias, some by higher powers of mass etc. (see \cite{Vlah:2012ni} for details). 
In the case of halos we concluded above that 
our modeling gives a fairly good estimate of the velocity dispersion for all correlators, meaning that the small scale 
velocity dispersion is small and the velocity correlators are dominated by large scale velocity flows.
However, in case of halos this procedure might be formally justified, but this only partially resums the series, leaving all 
of the other terms that do not contain velocity correlators untouched. Since velocity dispersion is much smaller in case of halos
than it is for dark matter, terms that do not contain velocity dispersion are more 
relevant for halos in a relative sense. Since resummation is effecting only velocity dispersion terms one should not 
expect that this will then dramatically improve the overall performance of RSD model. 

It is customary to expand the redshift-space power spectrum in terms of Legendre multipole moments
\begin{equation}
  P^{ss}(k,\mu)=\sum_{l=0,2,4,\cdots}P^{ss}_l(k){\cal P}_l(\mu),
\end{equation}
where ${\cal  P}_l(\mu)$ are ordinary Legendre polynomials and multipole moments, $P^{ss}_l$, are given by 
\begin{equation}
  P^{ss}_l(k)=(2l+1)\int^{1}_{0}P^{ss}(k,\mu){\cal P}_l(\mu)d\mu ~. 
\end{equation}
In the RSD analysis it is common to model the monopole ($l=0$) and quadrupole ($l=2$) terms, since these contain most 
of the information on the angular structure of the correlations, 
although some information is also contained in the hexadecapole term ($l=4$), which we will not include here since 
it is dominated by $\mu^6$ terms that we do not explicitly model (although is present in our model through the resummation term).

\begin{figure}[t!]
    \centering
    \includegraphics[scale=0.295]{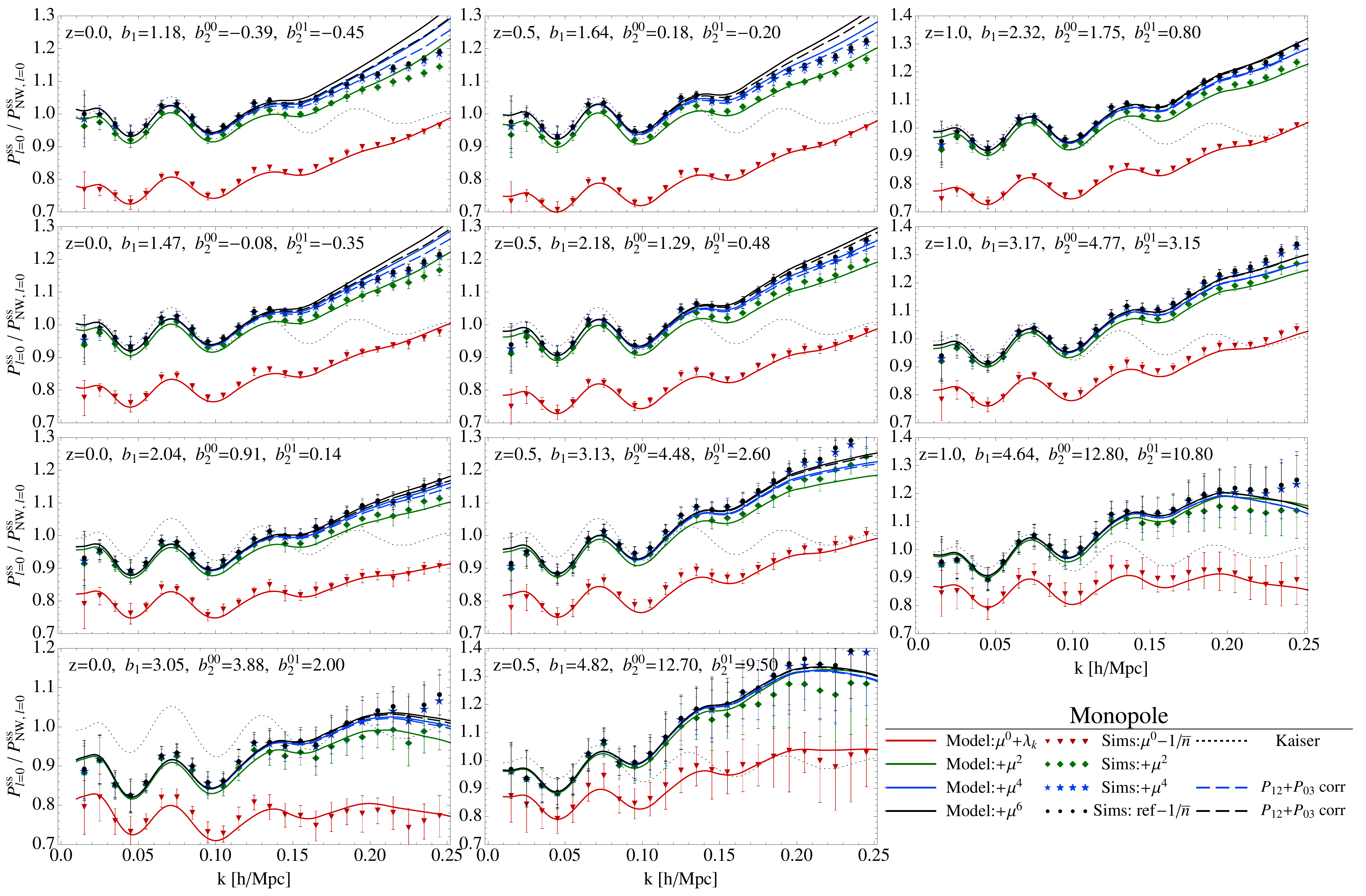}
    \caption{\small Monopole ($l=0$) shown for 
    several mass bins at redshifts $z=0.0$, $z=0.5$ and $z=1.0$. First the isotropic $P_{00}$ part (in red) is shown and 
    then we add the contributions of $\mu^2$ (green), $\mu^4$ (blue) and $\mu^6$ (black) part. Solid lines show
    the model presented in this paper and corresponding point marks simulation measurements of the same quantities. 
    We also show the model (dashed lines) when the correction to $P_{12}+P_{03}$ term
    is added to the $\mu^4$ term, as discussed in section \ref{subsec:1203}.  Direct simulation measurements of monopole (black points)
    is also shown. All the lines and data are divided by the Kaiser no-wiggle monopole prediction.}
    \label{fig:13}
\end{figure}

\begin{figure}[t!]
    \centering
    \includegraphics[scale=0.295]{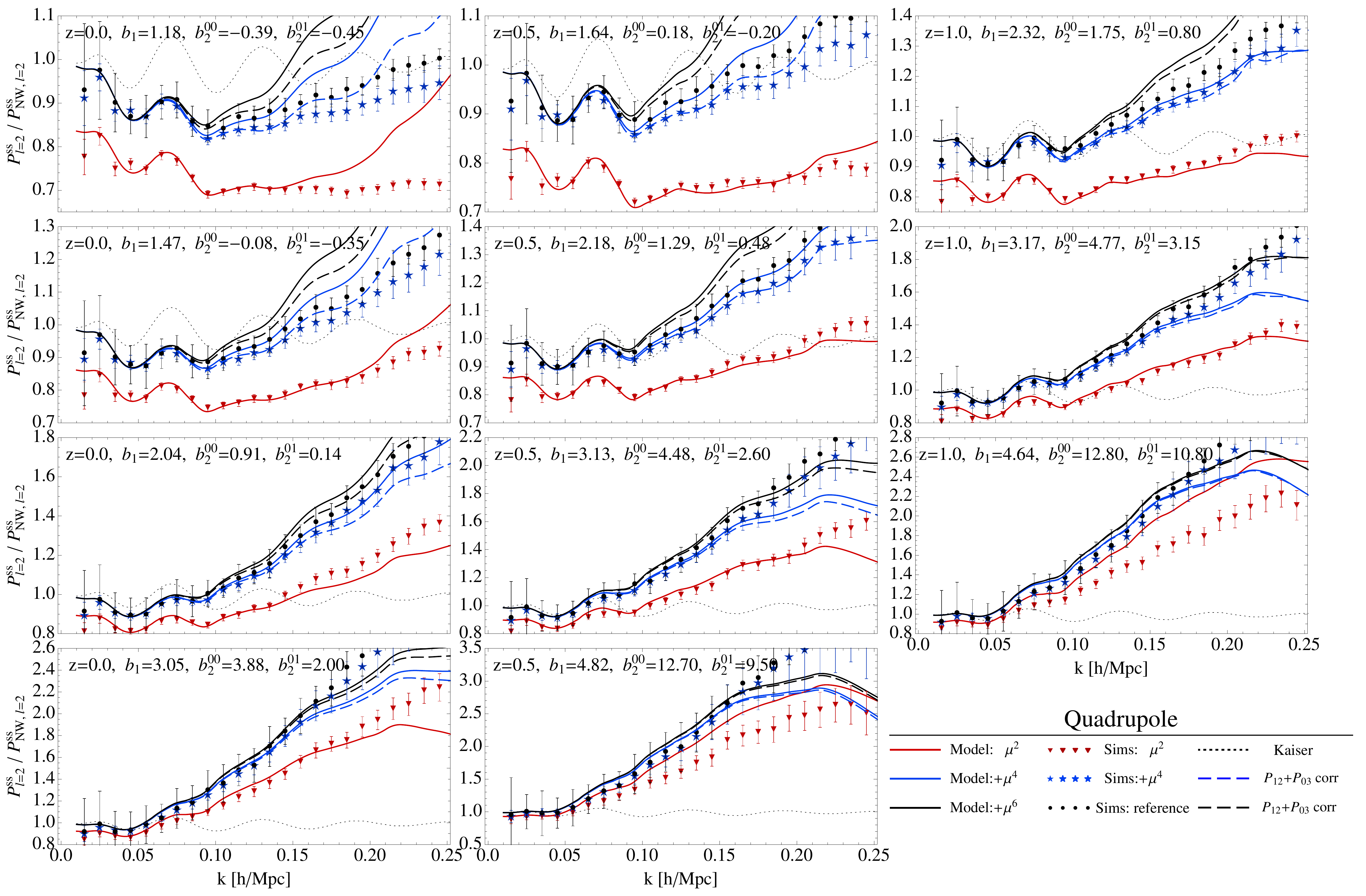}
    \caption{\small Quadrupole ($l=2$) shown for 
    several mass bins at redshifts $z=0.0$, $z=0.5$ and $z=1.0$. We show $\mu^2$ part (in red) and 
    then add the contributions of $\mu^4$ (blue) and $\mu^6$ (black) part. Solid lines show
    the model presented in this paper and corresponding point marks simulation measurements of the same quantities. 
    We also show the model (dashed lines) when the correction to $P_{12}+P_{03}$ term
    is added to the $\mu^4$ term, as discussed in section \ref{subsec:1203}.  Direct simulation measurements of monopole (black points)
    is also shown.
    All the lines and data are divided by the Kaiser no-wiggle quadrupole prediction.}
    \label{fig:14}
\end{figure}

In figures \ref{fig:13} and \ref{fig:14} we show monopole and quadrupole power spectra predictions. We show contributions 
to the multipoles as powers of $\mu$ and compare all the results to the reference multipole data obtained from full simulation redshift space power spectra. 
We also show simulation results where only terms up to $\mu^4$ are considered. In the case of monopole we see that these two simulation results 
agree on scales larger than $k\sim (0.10-0.15)$h/Mpc (depending on redshift and bias), but then start to deviate one from the other. 
These is especially apparent the case of the quadrupole, where we clearly see the difference in power when terms up to $\mu^4$ are considered, 
and the power when higher powers of $\mu$ are also taken into account, i.e.  terms proportional to $\mu^6$ and higher. 
We also show the result where one loop SPT prediction of $\mu^6$ term is added to the
model. We see that better modelling of also these $\mu^6$ terms is necessary to achieve more precise results in total.

\begin{figure}[t!]
    \centering
    \includegraphics[scale=0.292]{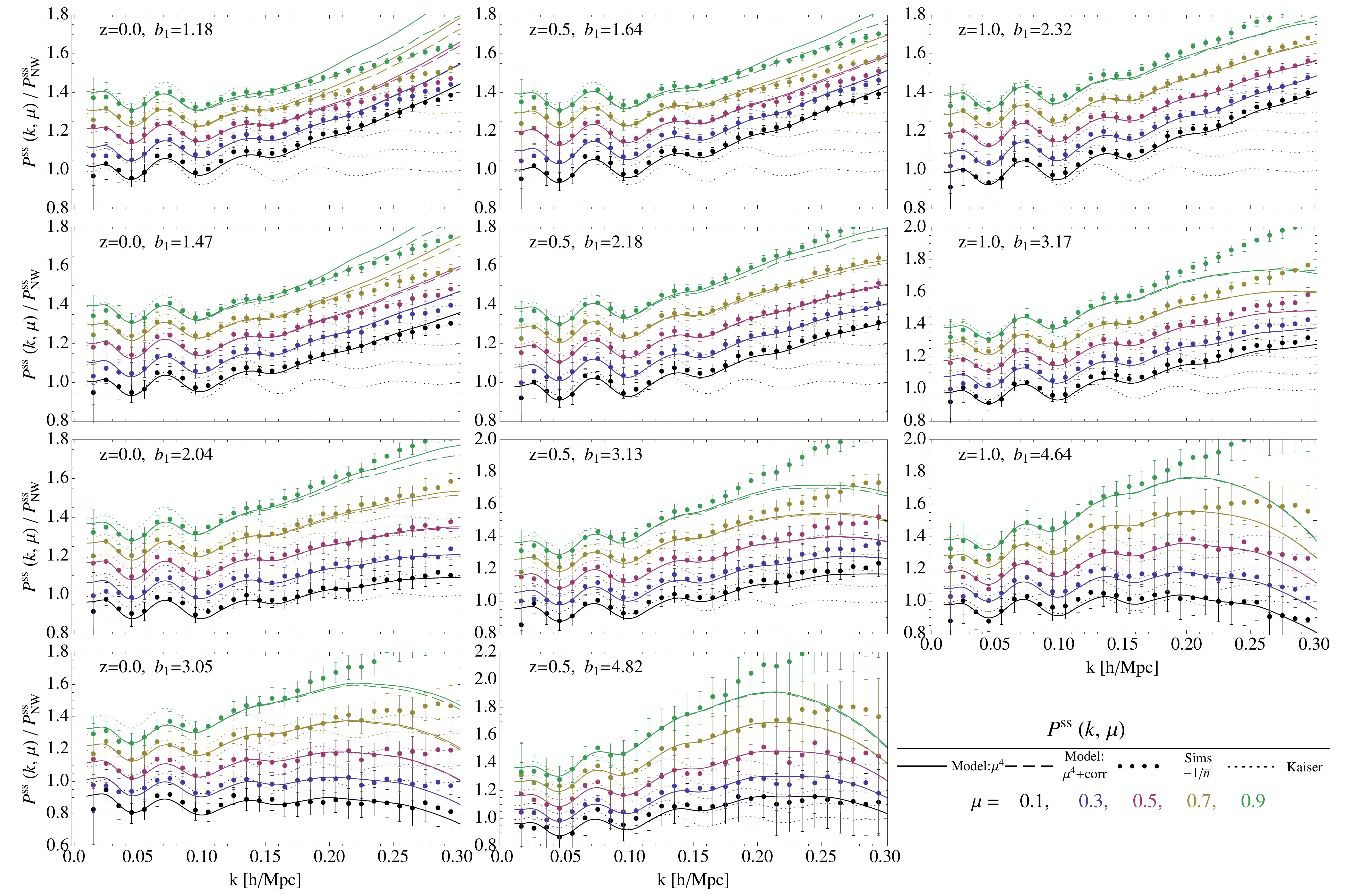}
    \caption{\small Angular dependence of the RSD model for 
    several mass bins at redshifts $z=0.0$, $z=0.5$ and $z=1.0$. We show the RSD model up to $\mu^4$ (solid lines),
    and $\mu^6$ (dashed lines). Simulation measurements (points) are also shown for each $\mu$ bin.
    We also show the model when the correction to $P_{12}+P_{03}$ term
    is added to the $\mu^4$ term (dashed line), as discussed in section \ref{subsec:1203}.
    All the lines and data are divided by the Kaiser no-wiggle predictions. Results for each angle bin 
    are offset for a constant value for a better overview.}
    \label{fig:15}
\end{figure}

In figure \ref{fig:15} we show the angular dependence of model versus simulations for five angular bins, also known in the literature as clustering wedges. 
Similar techniques have recently been used in analysis of the correlation function (e.g. \cite{Kazin:2013rxa}). We show the model up to $\mu^4$, with and 
without the correction on $P_{03}+P_{12}$ model and compare the results to the simulation measurements. We see that, as expected, the model is
better for the case of low $\mu$ since the expansion parameter of the distribution function approach is $k_\pp v$ and by construction 
we have a very good model for real space power spectrum.

%===============================================================%
%===============================================================%
\section{Correlation function}
\label{sec:cf}
%===============================================================%
%===============================================================%

Our model was built in Fourier space, but we can also look at its performance in configuration space. To get the correlation function we Fourier transform
the redshift space power spectra
\begin{align}
\xi^{ss}(\VEC{s})&=\int{\frac{d^3q}{(2\pi)^3}~P^{ss}(\VEC{q})e^{-i\VEC{q}\cdot\VEC{s}}}\nonumber\\
&=\xi_0(s){\cal P}_0(\nu)+\xi_2(s){\cal P}_2(\nu)+\xi_4(s){\cal P}_4(\nu)
\end{align}
where we used first four ordinary Legendre polynomials, ${\cal P}_0(\nu)=1$, 
${\cal P}_2(\nu)=(3\nu^2-1)/2$ and ${\cal P}_4(\nu)=(35\nu^4-30\nu^2+3)/8$, and 
$\nu$ is the cosine of the angle between $\VEC{s}$ and line of sight. Expansion coefficients are given by spherical 
Bessel function $j_l$ moments of the power spectra
\begin{align}
\xi_{l}(s)=i^l\int{\frac{q^2dq}{2\pi^2}~W_R(q)P^{ss}_l(q)j_l(qs)}.
\end{align} 

\begin{figure}[t!]
    \centering
    \includegraphics[scale=0.47]{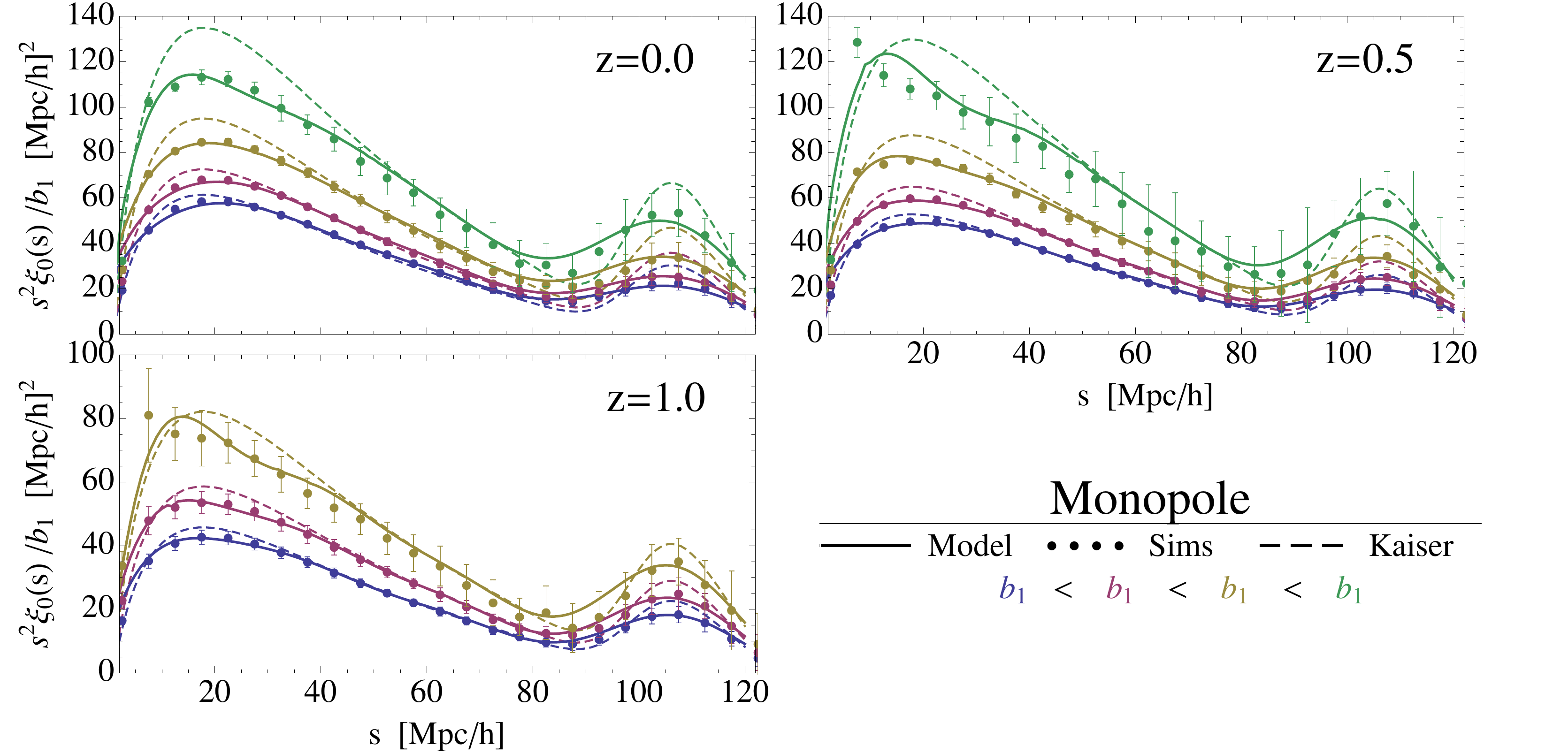}
    \caption{\small Monopole of redshift-space correlation function for several mass bins at redshifts $z=0.0$, $z=0.5$ and $z=1.0$.
    We show the results of model presented in this work (solid lines) linear theory predictions (dashed lines) and halo simulation
    measurements (points). Results are shown for the same mass bins as in previous plots 
(upper line represents the highest bias and lower line represents the lowest bias, respectively).}
    \label{fig:16}
\end{figure}

\begin{figure}[t!]
    \centering
    \includegraphics[scale=0.47]{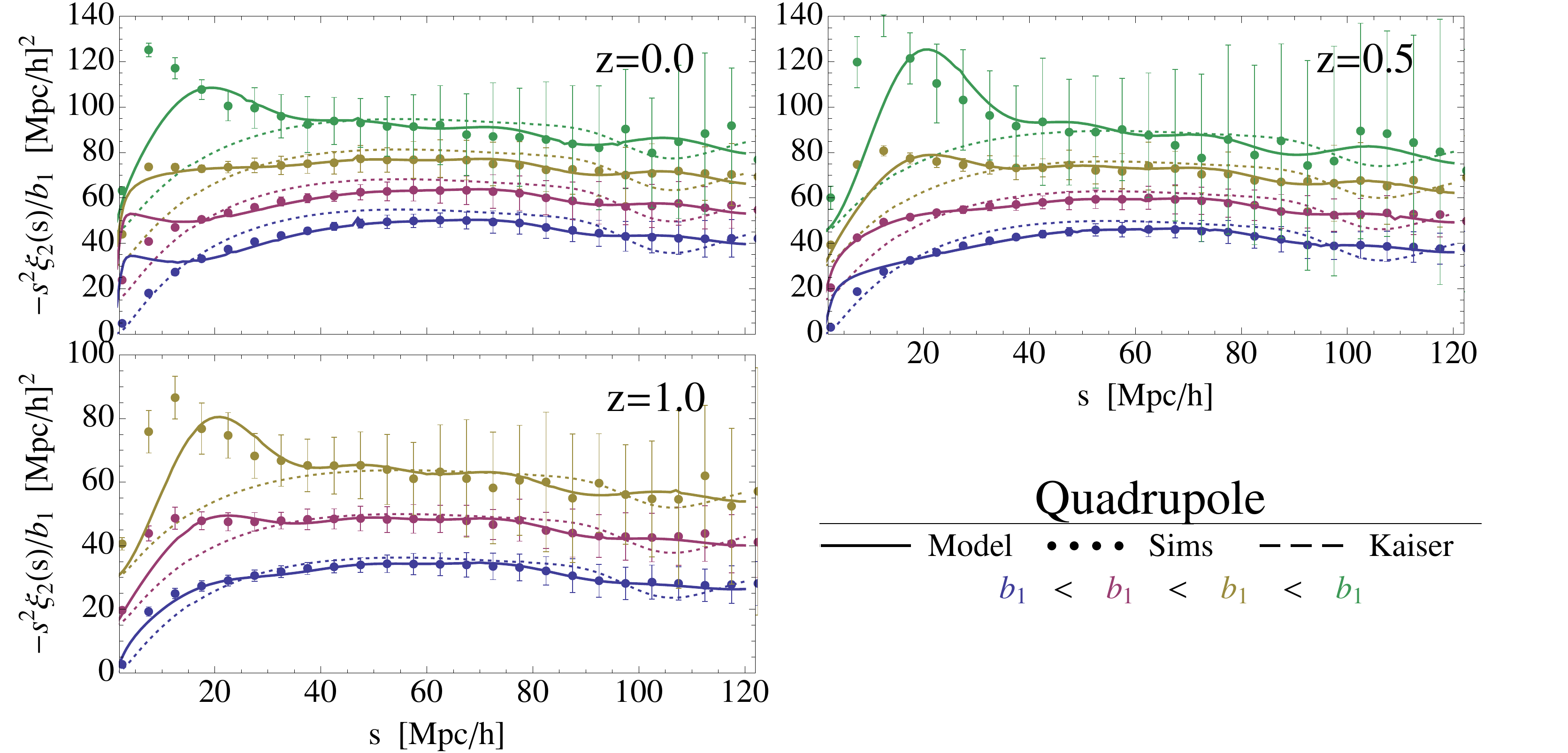}
    \caption{\small Same as figure \ref{fig:16} but for quadrupole moment of redshift-space correlation function. 
    Results for each mass bin are offset by a constant value (15 $(\text{Mpc/h})^2$) relative to the lowest (blue) bias.}
    \label{fig:17}
\end{figure}

However, many of our PT model predictions strongly diverge from simulations at high $k$, a well known problem of PT. 
To cure this we introduce the window function $W(qR)$ with smoothing radius $R$, which reduces the importance of high $k$ 
contributions to the correlation function. For the smoothing $W(qR)$ function we use the simple Gaussian filter. 
This suppresses the amplitude of the correlation function, and the effect is stronger as we approach smaller scales.
We choose the value for which the filter effects on scales larger than $s=5Mpc/h$ are small and 
are not noticeable in the figures \ref{fig:16} and \ref{fig:17} presented bellow. In principle stricter criteria could be implemented here 
to quantify these effects, but for our purposes this is not of the crucial importance. 
We find the value to be $R=1.0h/Mpc$ for both monopole and quadrupole case.
In figures \ref{fig:16} and \ref{fig:17} we show monopole and quadrupole predictions in
configuration space, obtained by Fourier transforming the model presented in this work. We show the results for several mass bins and 
redshifts and compare them to the N-body simulation measurements. 
We note that the model is not supposed to be compared against simulations below $s=10Mpc/h$ because of the artificial smoothing 
of the model against simulations. In principle we could have inserted the smoothing also into the simulations,
but for this procedure a broad range of scales of the correlation function measurements is needed.

%===============================================================%
\section{Conclusions}
\label{sec:conclusion}
%===============================================================%

In this paper we continue the studies of distribution function approach to redshift space distortions (RSD), applying Eulerian perturbation 
theory (PT) to the case of dark matter halos. 
In this formalism the RSD power spectrum is decomposed into moments of distribution function and  our goal is to model 
these contributing terms using the  perturbation theory and Eulerian biasing model.
We work at 1-loop level in PT, requiring us to introduce 3 additional biasing parameters, of which we find that one, the tidal 
tensor bias $b_s$, turns out not to be important if one assumes its amplitude is given by the bispectrum analysis \cite{Baldauf:2012hs, Chan:2012jj}. 
The remaining two nonlinear bias parameters are the local quadratic bias $b_2$ and the non-local 3rd order bias $b_{3,nl}$. 
These two have similar scale dependence with $k$, at least over a limited range of $k<0.15h/Mpc$, but have different amplitudes in the 
density-density correlator $P^{hm}_{00}$ relative to the density-momentum correlator $P^{(hh)}_{01}$. One can thus parametrize these nonlinear biases with 2 
independent effective 2nd order bias terms. Simple coevolution theory predicts that the nonlinear biasing should be weaker in density-momentum relative to density-density 
\cite{Saito:2013}, and our results confirm this prediction. 

We require that our biasing scheme is consistent with other statistics, in particular
the halo-dark matter density cross-correlation $P^{hm}_{00}$ \cite{Saito:2013}. We also assume that the 
tidal tensor bias $b_s$ is consistent with the bispectrum \cite{Chan:2012jx,Baldauf:2012hs}. 
We thus start by modelling the halo matter cross correlation
term $P^{(hm)}_{00}$ where we fit for effective $b^{00}_2$ values, combining all the nonlinear bias terms into one. 
However, since the statistic that enters the RSD is the halo-halo density correlator $P_{00}^{hh}$, this means we also 
need to describe the stochasticity $\Lambda(k)=P_{00}^{hh}-2b_1P_{00}^{hm}+b_1^2P_{00}^{mm}$. 
Detailed modeling of this term is complicated, and is related to halo exclusion and 
nonlinear biasing \cite{Baldauf:2013}. Since our goal is to 
study RSD we do not attempt to develop a more detailed model of this term
and instead we simply parametrize it with a simple power law expression. 
For the most of the mass bins that were considered we encounter sub-Poissonian stochasticity. 
We next turn to the modelling of higher momentum correlations, for which the nonlinear contributions of bias at one loop level enter 
explicitly only in the $P^{(hh)}_{01}$ term. We find the values of effective 
$b^{01}_2$ that reproduce the simulation measurements. We have argued that at the level of 1-loop calculations this approach is consistent, 
as we have both quadratic local bias $b_2$ and cubic non-local bias $b_{3,nl}$ entering at the same order, but with differing 
coefficients in $P_{00}^{hh}$ versus $P_{01}^{hh}$. 

In addition to biasing PT approach also computes dark matter clustering. However, 
in previous work \cite{Vlah:2012ni} we found that PT does not do a good job in predicting the dark matter correlators: this is 
a well known property of PT for density-density and density-velocity correlations (e.g. \cite{Carlson:2009it}), which also holds
for higher order density weighted velocity moment correlators. As a result we use the dark matter correlators as 
presented in \cite{Vlah:2012ni}, which were a combination
of simulation measurements and perturbation theory calculations. We divide the halo correlators into the dark matter part and the 
remaining part, which depends on linear and nonlinear bias parameters. We use PT to compute both linear and 
nonlinear biasing contributions to the halo correlators. For the dark matter part, we use 
$P_{00}$, $P_{01}$, $P_{11}$ and $P_{\delta \theta}$ as given by the dark matter simulations. Ideally this part will also be 
eventually replaced by PT, but since our goal here is not the dark matter modeling but the halo modeling we do not investigate it further. 

The dominant term to RSD is the $\mu^2$ term and its dominant contribution is 
the halo momentum density correlated with the halo density. Two other terms contribute to 
$\mu^2$, the vector part of the halo momentum density- halo momentum density correlation $P_{11}^{hh}$, and
the scalar part of halo kinetic energy density - halo density correlation $P_{02}^{hh}$. We find that they affect RSD at a 10\% level already at 
$k \sim 0.05 {\rm h/Mpc}$ for most of the mass bins we consider. 
The halo kinetic energy density- halo density correlation term $P_{02}$ is the dominant nonlinear effect, and is negative at all $k$
and thus reduces the total $\mu^2$ power. It is related to the Fingers-of-God (FoG) effect, since this term contains velocity
dispersion term. However, unlike the velocity dispersion inside the halos, which dominates FoG for dark matter,
this effect is generated by the large scale velocity flows which cancel out with $P_{11}$ term on small scales. 
As a result there are no velocity dispersion effects on small scales. 

The next angular term has $\mu^4$ dependence and there are 
seven terms that contribute to the total power spectrum, of which one, 
scalar part of $P^{(hh)}_{11}$, contains a linear order contribution that does not
vanish on large scales. Modelling these terms has proven to be even more difficult than lower order terms, 
but certain level of success has been achieved  compared to the simulations over a
limited dynamic range, with errors of about 10\% at $k \sim 0.1 {\rm h/Mpc}$ at $z=0$. 
All these terms exhibit very strong scale dependence, which we can reproduce in our model, 
which should be viewed as a success since there are no free parameters used. 
Nevertheless, some of the terms also have a considerable error, specially in $P_{03}^{hh}+P_{12}^{hh}$, and as a result we
do not succeed in modeling accurately the modes above $k \sim 0.15h/Mpc$. 

Our ultimate goal is to develop an accurate RSD model that can be applied to observations, but this was not the 
primary focus of this work. Instead, here we focused on asking whether we can model all the diferent halo density 
weighted powers of velocity using a consistent halo biasing model at 1-loop level. 
We emphasize that all of our biasing parameters are physically motivated: indeed, in most cases they can be predicted
from a biasing model \cite{Baldauf:2012hs,Chan:2012jj,Saito:2013}
and all of the bias parameters exhibit a simple halo mass dependence that can be used as a prior when applying these models 
to the real data. We introduced no arbitrary velocity dispersion parameters, like those needed
in models of most previous work on the subject \cite{Scoccimarro:2004tg,Taruya:2013my}. 
This is because when it comes to halos there is no small scale velocity dispersion, as
the halos centers are at rest with respect to the local center of mass. 
All velocity dispersion effects come from large scale velocities which are fully modeled in our approach using PT. We do not compute higher 
order velocity effects beyond $\mu^4$ and instead we propose a simple resummation ansatz that should approximately capture these terms. 
We have achieved some level of accuracy with our modeling, but 
a 1\% precision, needed for current and future RSD surveys,
can only be achieved up to $k \sim 0.15h/Mpc$. We have seen that the nonlinear effects at that $k$ are at 10-20\% level or larger and 
rapidly growing towards higher $k$, making it difficult to significantly improve the model beyond what was achieved here. 
Performance of the presented RSD model in determining the cosmological parameter and comparison to some of the other models
will be studied in \cite{Blazek:2013}.

Successful modeling of halo velocity statistics is just one ingredient of the complete RSD model. 
We observe galaxies, not dark matter halos, and our analysis remains to be extended to galaxies. 
We saw previously that for dark matter we had to introduce small scale velocity dispersion to model FoG effects in RSD and 
we expect the same to be true for galaxies. 
Nevertheless, separating halo biasing effects, the focus of this work, from FoG effects is an important step towards the complete RSD model. 
The treatment of FoG we used for dark matter, based on the halo model
for computing velocity dispersion, should also be applicable to galaxies. 
We plan to address this in the future work.

%===============================================================%
\section*{Acknowledgments}
%===============================================================%

We would like to thank Patrick McDonald, Shun Saito, Tobias Baldauf, Jonathan Blazek, Masanori Sato, Jaiyul Yoo and Nico Hamaus for useful discussions and comments. 
ZV would like to thank the Berkeley Center for Cosmological Physics and the Lawrence Berkeley Laboratory for their hospitality.
This work is supported by the DOE, the Swiss National Foundation under contract 200021-116696/1 and WCU grant R32-10130. V.D. acknowledges support by the Swiss National Science Foundation.
The simulations were performed on the ZBOX3 supercomputer of the Institute for Theoretical Physics
at the University of Z\"{u}rich. For making some of the plots in this paper LevelScheme package \cite{LevelScheme} has been used.

\vfill

\bibliographystyle{JHEP}
\bibliography{Bib}

\end{document}